\catcode`\@=11
\newif\if@fewtab\@fewtabtrue
{\count255=\time\divide\count255 by 60
\xdef\hourmin{\number\count255}
\multiply\count255 by-60\advance\count255 by\time
\xdef\hourmin{\hourmin:\ifnum\count255<10 0\fi\the\count255}}
\def\ps@draft{\let\@mkboth\@gobbletwo
    \def\@oddhead{}
    \def\@oddfoot
       {\hbox to 7 cm{$\scriptstyle Draft\ version:\ \draftdate$
       \hfil}\hskip -7cm\hfil\rm\thepage \hfil}
    \def\@evenhead{}\let\@evenfoot\@oddfoot}

\def\ceqno{\global\@fewtabfalse
    \ifcase\@eqcnt \def\@tempa{& & &}\or \def\@tempa{& &}
      \or \def\@tempa{&}
      \or\def\@tempa{}\fi\@tempa
{\rm(\theequation)}}

\def\aeqno#1{\global\@fewtabfalse
    \ifcase\@eqcnt \def\@tempa{& & &}\or \def\@tempa{& &}
      \or \def\@tempa{&}
      \or\def\@tempa{}\fi\@tempa
{\rm(\theequation,#1)}}

\def\label#1{\ifnum\draftcontrol=1
 \global\def\draftnote{$\scriptstyle #1$}\fi
 \@bsphack\if@filesw {\let\thepage\relax
   \def\protect{\noexpand\noexpand\noexpand}%
\xdef\@gtempa{\write\@auxout{\string
      \newlabel{#1}{{\@currentlabel}{\thepage}}}}}\@gtempa
  \if@nobreak \ifvmode\nobreak\fi\fi\fi
  \@esphack}

\def\alabel#1#2{\label{#1}\global\@fewtabfalse
    \ifcase\@eqcnt \def\@tempa{& & &}\or \def\@tempa{& &}
      \or \def\@tempa{&}
      \or\def\@tempa{}\fi\@tempa
{\hbox to 3cm{\phantom{\rm(\theequation,#2)}
\draftnote \hfil}\hskip -3cm {\rm(\theequation,#2)}}}

\def\clabel#1{\label{#1}\global\@fewtabfalse
    \ifcase\@eqcnt \def\@tempa{& & &}\or \def\@tempa{& &}
      \or \def\@tempa{&}
      \or\def\@tempa{}\fi\@tempa
{\hbox to 3cm{\phantom{\rm(\theequation)}
\draftnote \hfil}\hskip -3cm{\rm(\theequation)}}}

\def\eqnarray{\def\draftnote{{}}\global\@fewtabtrue
\stepcounter{equation}\let\@currentlabel=\theequation
\global\@eqnswtrue
\global\@eqcnt\z@\tabskip\@centering\let\\=\@eqncr
$$\halign to \displaywidth\bgroup\@eqnsel\hskip\@centering\@eqcnt\z@
  $\displaystyle\tabskip\z@{##}$&\global\@eqcnt\@ne
  \hskip 1\arraycolsep \hfil${##}$\hfil
  &\global\@eqcnt\tw@ \hskip 1\arraycolsep
$\displaystyle\tabskip\z@{##}$
\hfil  \tabskip\@centering&\global\@eqcnt\thr@@\llap{##}\tabskip\z@
\cr}

\def\endeqnarray{\@@eqncr\egroup
      \global\advance\c@equation\m@ne$$\global\@ignoretrue}

\def\@eqnnum{\hbox to 3cm{\phantom{\rm(\theequation)} \draftnote
                         \hfil}\hskip -3cm {\rm(\theequation)}}

\def\@@eqncr{\let\@tempa\relax
    \ifcase\@eqcnt \def\@tempa{& & &}\or \def\@tempa{& &}
      \or \def\@tempa{&}
      \or\def\@tempa{}
\fi\@tempa
\if@eqnsw
\if@fewtab\@eqnnum\fi
\stepcounter{equation}\fi\global
\@eqnswtrue\global\@eqcnt\z@\global\@fewtabtrue\cr}

\def\draftcite#1{\ifnum\draftcontrol=1#1\else{}\fi}

\def\@lbibitem[#1]#2{\item{}\hskip -3cm \hbox to 2cm
{\hfil$\scriptstyle\draftcite{#2}$}\hskip
1cm[\@biblabel{#1}]\if@filesw
     {\def\protect##1{\string ##1\space}\immediate
      \write\@auxout{\string\bibcite{#2}{#1}}}\fi\ignorespaces}

\def\@bibitem#1{\item\hskip -3cm \hbox to 2cm
{\hfil $\scriptstyle\draftcite{#1}$}\hskip 1cm
\if@filesw \immediate\write\@auxout
       {\string\bibcite{#1}{\the\value{\@listctr}}}\fi\ignorespaces}

\def\draftdate{\number\month/\number\day/\number\year\ \ \ \hourmin }

\global\def\draftcontrol{0}
\catcode`\@=12

\catcode`\@=12

\documentstyle[11pt,epsf]{article}
\def\theequation{{\arabic{equation}}}
\newcommand{\be}{\begin{eqnarray}}
\newcommand{\en}{\end{eqnarray}\vs 0.5 cm}
\newcommand{\non}{\nonumber}
\newcommand{\no}{\noindent}

\newcommand{\Nu}{{\bf u}}

\newcommand{\Nk}{{\bf k}}

\newcommand{\Nj}{{\bf j}}
\newcommand{\Ns}{{\bf s}}

\newcommand{\Nna}{{\bf \nabla}}

\newcommand{\NR}{{{\bf R}}}
\newcommand{\NT}{{{\bf T}}}
\newcommand{\NZ}{{{\bf Z}}}
\newcommand{\NN}{{{\bf N}}}
\newcommand{\NK}{{{\bf K}}}
\newcommand{\Nn}{{{\bf n}}}
\newcommand{\Nm}{{{\bf m}}}
\newcommand{\qq}{\begin{eqnarray}}

\newcommand{\da}{\partial}

\newcommand{\qqq}{\end{eqnarray}}

\newcommand{\tr}{\hbox{tr}}

\newcommand{\CB}{{\cal B}}

\newcommand{\CD}{{\cal D}}

\newcommand{\CF}{{\cal F}}

\newcommand{\CI}{{\cal I}}

\newcommand{\CL}{{\cal L}}

\newcommand{\CR}{{\cal R}}

\newcommand{\hf}{{_1\over^2}}

\newcommand{\vs}{\vskip}

\begin{document}
\
\vskip 1.7cm
\begin{center}
{\large{\bf{Exponential Mixing of the 2D Stochastic Navier-Stokes
Dynamics
}\footnote{Research partially
supported by EC grant FMRX-CT98-0175 and by
ESF/PRODYN.}}}

\vskip 1.4cm
J.Bricmont\\ UCL, Physique Th\'eorique,
 B-1348, Louvain-la-Neuve,
Belgium\\
\vskip 0.3cm
 A.Kupiainen,
R.Lefevere \\ Helsinki University,
Department of Mathematics,\\ P.O.Box 4, Helsinki 00014,
Finland
\end{center}
\vskip 1.5cm
\date{ }

\vskip 1.3 cm

\begin{abstract}
\vskip 0.3cm

\noindent We consider the Navier-Stokes equation on a
two dimensional torus with a random force which is white noise
in time, and excites only
a finite number of  modes. The number of excited
modes depends on the viscosity $\nu$, and
 grows like  $\nu^{-3}$ when $\nu$
goes to zero.  We prove that this Markov process
has a  unique  invariant measure and is
 exponentially mixing in time.
\end{abstract}
\vs  1.6cm

\section{Introduction}

Homogenous isotropic turbulence is often mathematically
modelled by Navier Stokes equation subjected to an external
stochastic driving force which is stationary in space and time
and "large scale", which in particular means smooth in space.
The status of the existence and uniqueness of solutions
to the stochastic PDE parallels that of the deterministic one.
In particular, in two dimensions, it holds under very general conditions.

However, for  physical reasons, one is interested in the
existence, uniqueness and properties of the stationary state of the
resulting Markov process. While the existence of such a state
follows with soft methods \cite{vf}, uniqueness, i.e.
ergodic and mixing properties of the process has been
harder to establish. In a nonturbulent situation, i.e.
with a sufficiently rough forcing this was established
in \cite{fm} and for large viscosity in \cite{mat}.
The first result for a smooth forcing was by Kuksin
and Shirikyan \cite{ks} who considered a periodically kicked
system with bounded kicks. In particular they could
deal with the case where only a finite number of modes
 are excited by the noise
(the number depends both on the
viscosity and the size of the kicks). In \cite{bkl2},
we proved uniqueness and exponential mixing for such
a kicked system where the kicks have a  Gaussian distribution, but we required
that there be a nonzero noise for each mode. In this
paper, we extend that analysis to the case where only finitely many
modes are excited, and the forcing is
 white noise in time. An
essential ingredient in our analysis is the Lyapunov-Schmidt
type reduction introduced in \cite{ks}, that allows
to transform the original Markov process with infinite dimensional
state space to  a non-Markovian process with
finite dimensional state space. We apply standard
ideas of statistical mechanics (high temperature expansions)
to this process to deduce mixing properties of the dynamics.
While preparing this manuscript we received a
preliminary draft \cite{ems} that claims similar results,
using a somewhat more probabilistic approach. We thank
these authors for communicating us their ideas, some of which
helped us to simplify our arguments, especially in Section 8 below.

\vs 3mm

We consider the stochastic Navier-Stokes equation for the
velocity field ${u}(t,x)\in \NR^2$ defined on the torus $\NT=
(\NR/2\pi \NZ)^2$:
\qq
d{ u}+(({ u}\cdot\Nna){ u}-{\nu}\Nna^2{ u}+\Nna p)dt=d{ f}
\label{ns}
\qqq
where ${ f}(t,x)$ is a Wiener process with covariance
\qq
Ef_\alpha(t,x)f_\beta(t',y)= \min\{t,t'\}
C_{\alpha\beta}(x-y)
\label{F}
\qqq
and $C_{\alpha\beta}$ is a smooth function
satisfying $\sum_\alpha\da_\alpha C_{\alpha\beta}=0$.
Equation (\ref{ns}) is supplemented with
the incompressibility condition $\Nna\cdot{ u}=0=\Nna\cdot{ f}$,
and we will also assume that
the  averages over the torus vanish:
$\int_{\NT}{ u}(0,x)=0=\int_{\NT}{ f}(t,x)$, which imply
that $\int_{\NT}{ u}(t,x)=0$
 for all times $t$.

It is convenient to change to dimensionless variables so that $\nu$
becomes equal to one.  This is achieved by setting
$
{ u}(t,x)=\nu u'(\nu t, x).
$
Then $ u'$  satisfies
(\ref{ns}), (\ref{F}) with $\nu$ replaced
by $1$, and $C$ by
$$
C'=\nu^{-3}C.
$$
>From now on, we work with such variables
and drop the primes. The dimensionless control parameter
in the problem is the (rescaled) energy injection rate
$\hf\tr\, C'(0)$ ,
customarily written as
$
({\rm Re})^3$
where ${\rm Re}$ is  the
 Reynolds number:
$$
{\rm Re}=\epsilon^{_1\over^3}\nu^{-1},
$$
and $\epsilon= \hf\tr\, C(0)$ is the energy injection rate in the
original units (for explanations of the terminology see \cite{frisch}).

In two dimensions, the incompressibility condition can be
conveniently solved by expressing the velocity field
in terms of the vorticity $\omega=\da_1u_2-\da_2u_1$. First
(\ref{ns}) implies the transport equation
\qq
d\omega+(({ u}\cdot\Nna)\omega-\Nna^2\omega)dt=db,
\label{ve}
\qqq
where $b=\da_1f_2-\da_2f_1$ has the covariance
\qq
Eb(t, x)b(t', y)=\min\{t,t'\}(2\pi)^{-1}\gamma( x- y)
\non
\qqq
with $\gamma = -2\pi\nu^{-3}\Delta\tr C$.

Next, going to the Fourier transform, $\omega_
k(t)={_1\over^{2\pi}}\int_{\NT}
e^{i k\cdot x}\omega(t, x)d x$, with $ k\in\NZ^2$; we may
express $ u$ as  $ u_k=i{_{(-k_2,k_1)}\over^{k^{2}}}
\omega_k$, and write
the vorticity equation as
\qq
d\omega(t)=F(\omega(t))dt +db(t),
\label{ee}
\qqq
where the drift is given by
\qq
F(\omega)_k=-k^2\omega_k+{_1\over^{2\pi}}\sum_{l\in\NZ^2\backslash\{{
0},k\}}
{_{k_1l_2-l_1k_2}\over^{|l|^{2}}}
\omega_{k-l}
\omega_l
\label{drift}
\qqq
and $\{b_k\}$ are Brownian
motions
 with ${\bar b_k}=b_{-k}$ and
$$
Eb_k(t)b_l(t')=\min\{t,t'\}\delta_{k ,
-l}\,\gamma_k .
$$

The dimensionless control parameter for the vorticity equation is
\qq
R=\sum_{k\in\NZ^2}
\gamma_k=2\pi\gamma(0)
\label{s3}
\qqq
which is proportional to the $\omega$ injection rate,
and also to the third power of the
 Reynolds number. We will be interested
in the turbulent region $R\to\infty$; therefore,
we will always assume below, when it is convenient,
 that $R $ is sufficiently large.

For turbulence one is interested in the properties of stationary state
of
the stochastic equation (\ref{ee}) in the case of {\it smooth} forcing
(see \cite{bkl1} for some discussion of this issue) and,
ideally, one would like to consider the case
where one excites only a finite number of modes,
$$
\gamma_k\neq 0 \;,\; k^2\leq N,
$$
 with $N$ of  order of one. In this paper we assume
that $N$ scales as
\qq
N=\kappa R,
\label{kappa}
\qqq
with $\kappa$ an absolute constant fixed below.
We take all the other $
\gamma_k=0$, although this condition can easily be
relaxed. Let us denote the minimum of the covariance
by
$$
\rho=\min\{|\gamma_{k}|\;|\;|k|^2\leq N\}.
$$

Before stating our result, we need some definitions.
Let $P$ be the orthogonal projection in $H=L^2(\NT)$
to the subspace $H_s$ of functions having zero Fourier
components for $|k|^2> N$. We will write
$$
\omega=s+l
$$
with $s=P\omega$, $l=(1- P)\omega$
(respectively, the small $k$ and large $k$ parts of $\omega$).
Denote also by $H_l$ the complementary subspace (containing
the nonzero components of $l$).
$H$ is our probability space, equipped with $\CB$, the
 Borel
$\sigma$-algebra.

The stochastic equation (\ref{ee}) gives rise to a Markov
process $\omega(t)$ and we denote by
$P^t(\omega,E)$ the transition probability of this process.

Our main result is the

\vspace{3mm}

\no{\bf Theorem.} {\it The stochastic Navier-Stokes equation
} (\ref{ee}) {\it defines a  Markov process with state space
$(H, \CB)$ and  for all $R<\infty$, $\rho>0$ it has a unique
invariant measure $\mu$ there.  Moreover,
$\forall \omega\in H$, for all Borel sets $ E\in H_s$ and
for all bounded H\"older continuous functions $F$ on $H_l$, we have,
\qq
|\int P^t(\omega,d\omega')1_E(s')F(l')-\int\mu(d\omega')1_E(s')F(l'))
|\leq C(\omega)||F||_\alpha e^{-mt}
\label{con}
\qqq
where $m=m(R,\rho, \alpha)>0$, $||F||_\alpha$ is the  H\"older
norm of exponent $\alpha$, and $C(\omega)
$ is a.s. finite.}

\vskip 2mm

\no{\bf Remark 1.} In a previous paper \cite{bkl1} we
have shown that, with probability 1, the functions on the support of such
a measure as constructed here are real analytic. In particular
all correlation functions of the form
$$
\int\mu(d\omega)\prod_i\nabla^{n_i}u(x_i)
$$
exist.

\vskip 2mm

\no{\bf Remark 2.} The parameters in our problem are
$R$ and $\rho$. All constants that do not depend on them
will be generically denoted by $C$ or $c$.
Besides, we write $C(X, Y, Z)$ for a ``constant" depending only on $X, Y, Z$.
These constants
can vary from place to place, even in the same equation.

\vskip 2mm

We close this section by giving the outline of the proof 
and explain its connection to  ideas coming from
Statistical Mechanics.

Let us start by observing that,
if we neglect the nonlinear term in
(\ref{ee}-\ref{drift}), we expect $\|\omega\|$ to be of
order $R^{\hf}$, for typical realizations of the noise
($R^{\hf}$ is the typical size of the noise, and the
$-k^2
\omega_k$ term will dominate in eq. (\ref{ee})
for larger values of $\|\omega\|$). It turns out that
similar probabilistic estimates hold for the full
equation (\ref{ee}) as shown in Section 3. Now, if
$\|\omega\|$ is of size $R^{\hf}$, the $-k^2
\omega_k$ term will dominate the nonlinear
 term (which is roughly of size $\|\omega\|^2$) in eq.
(\ref{ee}), for $|k| \geq \kappa R^{\hf}$, and one can
expect that those modes (corresponding to $l$ above) will
behave somewhat like the solution of the heat equation
and, in particular, that they will converge to a
stationary state.

Thus, the first step is to express the $l$-modes in terms
of the $s$-modes at previous times. This is done in Section 2 and
produces a process for the $s$-modes that is no longer
Markovian but has an infinite memory. In Statistical
Mechanics, this would correspond to a system of unbounded
spins (the $s$-modes) with infinite range interactions,
with the added complications that, here, the measure is
not given in a Gibbsian form, but only through a Girsanov
formula, i.e. (\ref{s1}) below, and that time is
continuous. Hence, we  have to solve several problems: the
possibility that $\omega$ be atypically large, the long
range ``interactions", and finally, showing that a
version of the $s$-process with a suitable cutoff is
ergodic and mixing.

The large $\omega$ problem is treated in Section 3, using
probabilistic estimates developped in \cite{bkl1}, which,
in Statistical Mechanics, would be called stability
estimates. The infinite memory problem is treated in
Sections 4 and 5, which are inspired by the idea of
``high temperature expansion" in Statistical Mechanics,
namely writing the Gibbs measure or, here, the Girsanov
factor, as sum of products of factors having a finite
range memory and which become smaller as that range
increases. However, in the situation considered here,
carrying out this expansion requires a careful and non
standard partition of the phase space (explained in
Section 4). The problem is that, even though for typical
noise, hence for typical $\omega$'s, the $l$-modes depend
exponentially weakly on their past (see Section 2), thus
producing, typically,  ``interactions" that decay
exponentially fast, they may depend sensitively on their
past when the noise is large. In the language of Statistical
Mechanices, atypically large noise produces long range
correlations.

This problem of sensitive dependence is coupled to
the last problem, that of the convergence of the
$s$-process with finite memory to a stationary state.
We have to
get lower bounds on transition probabilities and we can
prove those (see Section 8) only when the $s$-modes
remain for a sufficiently long time in a suitable region
of the phase space; thus, if we did not control the
sensitive dependence, we would not be able to carry out
that last step. Finally, in Section 7, we prove the
bounds on our ``high temperature" expansion and, in
Section 6, we use that expansion to prove the Theorem.
Note that, because we deal with a stochastic process, we
never have to ``exponentiate" our expansion, unlike what
one would usually has to do in Statistical Mechanics
(i.e., the analogue of the partition function here equals
$1$). The choice of $\kappa$ in (\ref{kappa}) is
explained in Remark 2 of Section 4.

\section{Finite dimensional reduction}

We will use an idea of \cite{ks} to reduce the problem of the
study of a Markov process with infinite dimensional
state space to that of a non-Markovian process with
finite dimensional state space.

For this purpose, write the equation (\ref{ee}) for
the small and large components of $\omega$ separately:
\qq
ds(t)&=&PF(s(t)+l(t))dt +db(t)\label{seq}\\
{_d\over^{dt}}l(t)&=&(1-P)F(s(t)+l(t)) .
\label{leq}
\qqq
The idea of  \cite{ks} is to solve the $l$ equation
for a given function $s$, thereby defining $l(t)$ as a function
of the entire history of $s(t')$, $t'\leq t$. Then
the $s$ equation will have a drift with memory.
Let us fix some notation.
For a time interval $I$ we denote the
restriction of $\omega$ (or $s$, $l$ respectively) to $I$ by
$\omega(I)$, and use the boldface notation ${\bf s}(I)$,
to constrast it with $s(t)$, the value of $s$ at a single time.
 $\|\cdot\|$ will denote the
$L^2$ norm. In \cite{bkl1} it was proven that, for any $\tau<\infty$,
there exists a set $\CB_\tau$
of Brownian paths $b\in C([0,\tau],H_{s})$ of full measure
such that, for $b\in \CB_\tau$,
(\ref{ee}) has a unique solution with
$||\omega(t)||<\infty$, $||\nabla\omega(t)||<\infty$
for all $t$ (actually, $\omega(t)$ is  real analytic).
In particular, the projections $s$ and $l$ of this solution
are in $C([0,\tau],H_{s(l)})$ respectively.

On the other hand, let us denote, given any ${\bf s}\in C([0,\tau],H_s)$, the
solution - whose existence we will prove below -
of (\ref{leq}), with  initial condition
$l(0)$  by $l(t,{\bf s}([0,t]),l(0))$.
More generally, given initial data $l(t')$ at time $t'<\tau$ and
${\bf s}([t',\tau])$, the solution of (\ref{leq})
is denoted, for $\sigma \leq \tau$, by
$l(\sigma, {\bf s}([t',\sigma]),l(t'))$
and the corresponding $\omega$ by
$\omega(\sigma, {\bf s}([t',\sigma]),l(t'))$.
The existence and key properties of those functions are given by:

\vs 2mm

\no{\bf Proposition 1.} {\it Let $l(0)\in H_l$ and
$s\in C([0,\tau], H_s) $ . Then $l(\cdot,{\bf s}([0,t]),l(0))\in C([0,\tau], H_l)
\cap L^2([0,\tau], H^1_l)$, where $H^1_l= H_l\cap H^1$, and $H^1$ is
the first Sobolev space. In particular,
\qq
\sup_{t\in [0,\tau]}\| l (t, {\bf s}([0,t]),l(0)) \| \leq
C(R, \sup_{t\in [0,\tau]} \|s(t)\|, \|l(0)\|)
\label{l10}
\qqq
where the notation $C(R, \sup_{t\in [0,\tau]} \|s(t)\|, \|l(0)\|)$
is defined in  Remark 2, Section 1.
Moreover, given two initial conditions $l_1 , l_2$ and $t\leq \tau$
\qq
\| l (t, {\bf s}([0,t]),l_1) - l(t,{\bf s}([0,t]),l_2) \| \leq
\exp \left[- \kappa Rt
+ a\int^t_0 \| \Nna \omega_1 \|^2\right]
\| l_1-l_2\|
\label{21}
\qqq
where $a=(2\pi)^{-2}\sum |k|^{-4}$ and
$\omega_1(t)=s(t)+l_1(t,{\bf s}([0,t]),l_1)$. The solution satisfies
\qq
l(t, {\bf s}([0,t]),l(0))=l(t, {\bf s}([\tau,t]),l(\tau, {\bf s}([0,\tau]),l(0))).
\label{22}
\qqq
}

\vs 2mm

\no{\bf Proof}.  The existence of $l$ follows from standard
a priori estimates which we recall for completeness.
We have from (\ref{leq}) (see also (\ref{ve})), for sufficiently smooth
$l$,
$$
{1\over 2} {d \over dt} \| l \|^2  = - \| \Nna  l \|^2 +
(l, u \cdot \Nna s)
$$
since, by incompressibility, $\Nna\cdot u=0$, $(l, u \cdot \Nna l)=
\hf\int  \Nna \cdot(ul^2)=0$.
Use now the bound,
for the functions $d,v,b$,
\qq
|(d,v\cdot \Nna b)|\leq
\| d\| \| v \|_\infty \| \nabla b \|
\leq {\sqrt a} \| d \| \| \Delta v \| \| \Nna b\|
\label{nonlin}
\qqq
which follows from
$\| v \|_\infty\leq (2\pi)^{-1}\sum_k \frac{|v(k)|k^2}{k^2}$
and Schwarz' inequality, and where
$a=(2\pi)^{-2}\sum |k|^{-4}$.
Using (\ref{nonlin}), $\alpha \beta \leq \hf(\alpha^2 + \beta^2)$
and $\| \Delta u \|
=\| \Nna (s+l) \| $, we get:
\qq
|(l, u \cdot \Nna s)|&\leq& {\sqrt a} \|l\| (\|\Nna s \|+
\|\Nna l \|) \| \Nna s\|
\nonumber\\
&\leq& {_{\sqrt a}\over^2}(\|l\|^2+\|\Nna s \|^4)+\hf\|\Nna l \|^2
+{_{a}\over^2}\|l\|^2\| \Nna s\|^2.
\non
\qqq
Hence,
\qq
{d \over dt} \| l \|^2  \leq - \| \Nna  l \|^2 +
({\sqrt a}+a\| \Nna s\|^2)\|l\|^2+{\sqrt a}\|\Nna s \|^4.
\label{s4}
\qqq
The bound (\ref{l10}) on $||l(t)||$ follows then, by Gronwall's
inequality,
from (\ref{s4}) and
the finiteness of  $\sup_t\| \Nna s\|^2$ and of $\sup_t\|\Nna s \|^4$
(which follow from the finiteness of $\sup_t\| s\|^2$,
since $s$ has only finitely many nonzero
 Fourier
coefficients). Finally, the boundedness
of $\int_0^\tau||\Nna l||^2$ follows from (\ref{s4}) by integration.

For the second claim, let
$\delta l (t) = l (t,s,l_1)-l(t,s,l_2)\equiv l_1(t)-l_2(t)$, and
define $u^l=(1-P)u$. We have:
\qq
{1\over 2} {d \over dt} \| \delta l \|^2  = - \| \Nna \delta l \|^2 +
(\delta l,
\delta u^l \cdot \Nna \omega_1 + u_1 \cdot \Nna \delta l+
\delta u^l\cdot \Nna \delta l )
= - \| \Nna \delta l \|^2 + (\delta l, \delta
u^l \cdot \Nna \omega_1)
\label{a9}
\qqq
using, as above, $(\delta l, u_1 \cdot \Nna \delta l ) =0=
(\delta l,\delta u^l\cdot \Nna \delta l)$, and defining
$\omega_1 = s + l_1$. Now,
estimate, using (\ref{nonlin}) and $\| \Delta \delta u^l \|
=\| \Nna \delta l \| $,
\qq
|(\delta l,\delta u^l \cdot \Nna \omega_1)|
\leq {\sqrt a}  \| \delta l \| \| \Nna \delta l \| \| \Nna
\omega_1 \|
\leq \hf (\| \Nna \delta l \|^2 + { a}  \|
\delta l \|^2 \|\Nna \omega_1\|^2).
\label{a10}
\qqq
 So, by (\ref{kappa}) and the fact that
  $l_k\neq 0$ only for $k^2>N$,
\qq
{d \over dt} \| \delta l \|^2 \leq -
\kappa R \| \delta l \|^2 + { a}  \| \delta l \|^2 \|
\Nna \omega_1 \|^2,
\label{a11}
\qqq
which implies the claim (\ref{21}) using Gronwall's inequality.
The last claim
(\ref{22}) is obvious.\hfill $\Box$

\vs 4mm

Now, if $s=P\omega$ with $\omega$ as above being the solution
of (\ref{ee}) with noise $b\in \CB_\tau$ then the $l(s)$
constructed in the Proposition equals $(1-P)\omega$
and the stochastic process $s(t)$ satisfies the reduced equation
\qq
ds(t)= f(t)dt+db(t)
\label{sred}
\qqq
with
\qq
f(t)=
PF(\omega(t)).
\label{f}
\qqq
where $\omega(t)$ is the function on $C([0,t],H_s)\times H_l$
given by
\qq
\omega(t)=s(t)+l(t,{\bf s}([0,t]),l(0))
\label{oomega}
\qqq
(\ref{sred}) has almost surely bounded paths and we
have a
Girsanov representation for the transition probability
of the $\omega$-process in terms of the $s$-variables
\qq
P^t(\omega(0),F)
=\int \mu^t_{\omega(0)}(d\Ns)F(\omega(t))
\label{girs}
\qqq
with
\qq
\mu^t_{\omega(0)}(d\Ns)=e^{\int_0^t(f(\tau),\gamma^{-1}(ds(\tau)
-\hf f(\tau)d\tau))}\nu^t_{s(0)}(d\Ns)
\label{s1}
\qqq
where $\nu^t_{s(0)}$ is the Wiener measure with covariance
$\gamma$ on paths ${\bf s}={\bf s}([0,t])$ with starting point $s(0)$ and
$(\cdot,\cdot)$
the $\ell^2$ scalar product.  We define
the operator $\gamma^{-1}$ in terms of its action on the Fourier
coefficients:
\qq
(f,\gamma^{-1}f)=\sum_{|k|^2\leq N}|f_k|^2\gamma_k^{-1}.
\label{norga}
\qqq

\vs 3mm

The Girsanov representation (\ref{girs}) is convenient since
the problem of a stochastic PDE has been reduced to that of
a stochastic process with finite dimensional state space.
The drawback is that this process has infinite memory.
In Sections 4 and 5 we present a formalism, borrowed
from statistical mechanics, that allows us to approximate it by
a process with finite memory; the approximation will be controlled
in Section 7,  while the finite memory process will be studied in
Section 8. This analysis is mostly done in the $s$-picture, but
an important ingredient in it will be some a priori estimates
on the transition probabilities of the original Markov process
generated by (\ref{ee}) that we prove in the next Section.

\section{A priori estimates on the transition probabilities}

The memory in the process (\ref{sred}) is coming from
the dependence of the solution of (\ref{leq}) on its
initial conditions. By Proposition 1, the dependence
is weak if $\int_0^t\|\Nna\omega\|^2$ is less than
$cR$ for a suitable $c$. We localize the time intervals where this
condition holds by inserting a suitable partition of unity
in the expression (\ref{girs}). We shall show (in Section 8 below)
that, during such
time intervals,
the $s$ process  behaves qualitatively like an ergodic Markov
process. In this section we show that the complementary time
intervals occur with small probability.

Let us first explain the partition of unity.
We define, for each unit interval $ [n-1,n]\equiv \Nn$, a
quantity  measuring the
size of $\omega$ on that interval by:
\qq
D_n=
\hf\sup_{t\in \Nn}
||\omega(t)||^{2}+\int_{\Nn}||\nabla\omega(t)||^{2}dt.
\label{Dn}
\qqq
Let
$\{\phi_k\}_{k\in\bf N}$ be a smooth partition of unity
for $\NR^+$, with the support of $\phi_k$  contained in $[2^{k}R,2^{k+2}R]$
for $k>0$, and in $[0,4R]$
for $k=0$.
Set, for ${\bf k}\in\NN^{t}$,
\qq
\chi_{\bf k}(\omega)=\prod_n\phi_{k_n}(D_n(\omega)) .
\label{chik}
\qqq
We insert
$
1=\sum_{\bf k}\chi_{\bf k}
$
in (\ref{s1}), to get
\qq
\mu^t_{\omega(0)}(d\Ns)
=\sum_{\bf k}
\chi_{\bf k}\mu^t_{\omega(0)}(d\Ns).
\label{PT}
\qqq

The following Proposition bounds the probability of the
unlikely event that we are interested in:

\vs 3mm

\noindent
{\bf Proposition 2}. {\it There
exist constants $c > 0$, $c'<\infty$,
 $\beta_0 < \infty$, such that
for all $t,t'$, $1 \leq t < t'$ and all $\beta\geq \beta_0$,
\qq
P \Bigl(\sum^{t'-1}_{n=t} D_{ n}
(\omega) \geq \beta R|t'-t| \Big|
\omega(0) \Bigr)
\leq  \exp({_1\over^R}c'e^{-t} \|\omega(0)\|^2)
\exp
(-c\beta |t'-t|)
\label{x1}
\qqq}
\vs 3mm

In order to prove  Proposition 2, we need some Lemmas.
We will start with a probabilistic analogue of
the so-called  enstrophy balance:

\vs 5mm

\no{\bf Lemma 3.1.} {\it For all $\omega(0)\in L^2$, and all $t\geq 0$,
\qq
E\Bigl[e^{{_{1}\over^{4R}}\|\omega(t)\|^2 }\;\Big|\;\omega(0)\Bigr]\leq
3e^{{_{1}\over^{4R}}e^{-t}\|\omega(0)\|^2},
\label{c4}
\qqq
and}
\qq
P( \|\omega(t)\|^2\geq D |\omega(0)) \leq 3
e^{-\frac{D}{4R}}
e^{{_{1}\over^{4R}}e^{-t}\|\omega(0)\|^2}
\label{c5}
\qqq

\vs 3mm

\noindent
{\bf Remark}. This Lemma shows that the distribution of
$\|\omega (t) \|^2$ satisfies an exponential bound on
scale $R$ with a prefactor whose dependence on the
initial condition decays exponentially in time. Thus, if
$\|\omega(0)\|^2$ is of order $D, \|\omega (t) \|^2$
will be, with large probability, of order $R$ after a
time of order $\log D$.

\vs 3mm

\no{\bf Proof.} Let $x(\tau)= \lambda(\tau)\|\omega(\tau)\|^2 =
\lambda(\tau) \sum_k
|\omega_k|^2$ for $0\leq \tau\leq t $. Then by Ito's
formula (remember that, by (\ref{s3}), $\sum_k\gamma_k =
R$ and thus $\gamma_k\leq R$, $\forall k$):
\qq
{_d\over^{d\tau}}E[e^x]&=&
E[(\dot\lambda\lambda^{-1}x-2\lambda\sum_k k^2
|\omega_k|^2+{\lambda}
\sum_k\gamma_k+{2\lambda^2}\sum_k
\gamma_k|\omega_k|^2)e^x]\nonumber \\
&\leq& E[((\dot\lambda\lambda^{-1}-2+2\lambda
R)x+\lambda R)e^x]
\label{c6}
\qqq
where  $E$ denotes the conditional expectation, given $\omega(0)$,
and where we used the
Navier-Stokes
equation (\ref{ve}), $|k|\geq 1$ for $\omega_k\neq 0$, and the fact that the
nonlinear term does not contribute (using integration by parts
and $\Nna\cdot u=0$). Take now
$\lambda(\tau)= {_1\over^{4R}}e^{(\tau-t)}$ so that $\lambda \leq
{_1\over^{4R}}$, $\dot\lambda\lambda^{-1}= 1$,
$\dot\lambda\lambda^{-1}-2+2\lambda
R
\leq -\frac{1}{2}$ and $\lambda R\leq  \frac{1}{4}$. So,
\qq
{_d\over^{d\tau}}E[e^x]\leq E[(\frac{1}{4}-\frac{1}{2}x)e^x]\leq
\frac{1}{2}-\frac{1}{4}E[e^x]
\non
\qqq
where the last inequality follows by using
$(1-2x)e^x\leq
2-e^x$. Thus,  Gronwall's
inequality implies that:
\qq
E[e^{x(\tau)}]\leq e^{-\frac{\tau}{4}}e^{x(0)} +2\leq 3e^{x(0)}
\non
\qqq
i.e., using the definition of $\lambda(\tau)$,
\qq
E\Bigl[\exp(\frac{e^{\tau-t}}{4R}\|\omega(\tau)\|^2 )\Bigr]\leq
3\exp(\frac{e^{-t}\|\omega(0)\|^2}{4R}),
\non
\qqq
This proves (\ref{c4}) by putting $\tau=t$;
(\ref{c5}) follows from (\ref{c4}) by Chebychev's
inequality. \hfill$\Box$

\vs 5mm

Since the $D_n$ in (\ref{x1}) is
the supremum over unit time intervals of
\qq
D_t(\omega) = {1\over 2} \| \omega (t) \|^2 + \int^t_{n-1}
\|
\Nna \omega \|^2 d \tau \hspace{5mm} n-1 \leq t \leq n,
\label{a0}
\qqq
which does not involve only
 $ \| \omega (t) \|^2$, we need
to control also the evolution of $D_t(\omega)$ over a
unit time interval, taken, for now,
to be $[0,1]$. From the Navier-Stokes equation
(\ref{ve}) and Ito's formula, we obtain
\qq
D_t(\omega) = D_0 (\omega) + Rt + \int^t_0 (\omega, db)
\label{a01}
\qqq
(since the nonlinear term does not contribute, as in (\ref{c6})).
Our basic estimate is:

\vs 3mm

\no{\bf Lemma 3.2.} {\it There exist $C < \infty$, $c>0$ such that,
$\forall A\geq 3D_0(\omega)$
\qq
P(\sup_{t\in [0,1]} D_t (\omega) \geq  A|\omega(0) )
\leq C e^{-{_{cA}\over^{R}}  }
\label{a1}
\qqq
}
\vs 3mm

\noindent
{\bf Remark}. While the previous Lemma showed that
$\|\omega (t) \|^2$ tends to decrease as long as it is
larger than ${\cal O} (R)$, this Lemma shows that, in a
unit interval, $D_t (\omega)$ does not increase too much
relative to $D_0 (\omega) = {1\over 2} \|\omega
(0)\|^2$. Thus, by combining these two Lemmas, we see
that $D_n (\omega)=\displaystyle{\sup_{t\in[n-1,n]}} D_t (\omega)$ is,
with large probability, less than $\|\omega (0)\|^2$,
when the latter is larger than ${\cal O} (R)$, at least
for $n\geq n_0$ not too small.
This is the content of Lemma 3.3. below.
Thus, it is unlikely that
$D_n (\omega)$ remains much larger than $R$ over some
interval of (integer) times, and this fact will be the basis of the
proof of Proposition 2.

\vs 3mm

\no{\bf Proof.} From (\ref{a01}), we get that
\qq
P\left(\sup_{t\in [0,1]} D_t (\omega) \geq A
\Big|\omega(0) \right)
\leq P\left(\sup_{t\in [0,1]} |
\int^t_0 (\omega,
db)| \geq (A - D_0 - R)\Big|\omega(0)\right).
\label{a2}
\qqq
The process $t\rightarrow \int^t_0 (\omega, db)$ is a continuous
martingale so, by Doob's inequality (see e.g.\cite{simon}, p.24), the submartingale
$x_t \equiv | \int^t_0 (\omega, db)|$ satisfies the bounds
\qq
 E((\sup_t x_t)^p )\leq ( {_p\over^{ p-1}})^p E(x^p_1)
\;\;\forall p \geq 2,
\label{ax3}
\qqq
where  $E$ denotes the conditional expectation, given $\omega(0)$.
These imply
\qq
 E(e^{\varepsilon \sup x_t}) \leq 5 E (e^{\varepsilon x_1}),
\label{a3}
\qqq
where $\varepsilon$ will be chosen small below (to derive (\ref{a3}),
expand both exponentials, use (\ref{ax3}) and
$( {_p\over^{ p-1}})^p\leq 4$ for $p\geq 2$;
for $p=1$, use $Ea\leq \hf (\alpha + \alpha^{-1}Ea^2)$
for $a\geq 0$ and take $\alpha=2$). Since
\qq
E(e^{\varepsilon x_1}) \leq {1\over 2} \left(E(e^{\varepsilon
\int^1_0 (\omega,
db)}) + E(e^{-\varepsilon \int^1_0 (\omega, db)})\right),
\label{a4}
\qqq
using Novikov's bound, we get
\qq
&& E(e^{\pm \varepsilon \int^1_0 (\omega, db)})
 \leq \left( E
(e^{2\varepsilon^2\int^1_0 d\tau(\omega(\tau),\gamma
\omega(\tau))})\right)^{1/2}\non\\
&\leq& \left(\int^1_0 d\tau E
(e^{2\varepsilon^2 (\omega(\tau),\gamma
\omega(\tau))})\right)^{1/2} \leq \left(\int^1_0 d\tau E
(e^{2\varepsilon^2 R \| \omega (\tau)\|^2})\right)^{1/2},
\label{a04}
\qqq
where the last two inequalities follow from Jensen's inequality,
applied to $e^{2\varepsilon^2\int^1_0 d\tau(\omega(\tau),\gamma
\omega(\tau))}$, and  from $\gamma_k \leq  R$ (see (\ref{s3})).

So, altogether, we have, by Chebychev's inequality and
(\ref{a3}-\ref{a04}):
\qq
P\Bigl(\sup_{t\in [0,1]} |
\int^t_0 (\omega,
db) | \geq (A - D_0 - R)\Big|\omega(0)\Bigr) \leq
5e^{-\varepsilon(A - D_0-R)}
\left(\int^1_0 d\tau E
(e^{2\varepsilon^2 R \| \omega (\tau)\|^2})\right)^{1/2}
\qqq

Now, combine this with (\ref{a2}) and (\ref{c4}) in
Lemma 3.1 above, choosing $2\varepsilon^2 R = {1 \over
4R}$, i.e.
$
\varepsilon = {1\over \sqrt 8 R}$, to get
\qq
P(\sup_{t\in [0,1]} D_t(\omega) \geq A
|\omega(0)) \leq 15 e^{-\varepsilon(A -
D_0-R)}e^{{D_0/ 4R}}
\label{a6}
\qqq
which yields (\ref{a1})
for $A \geq 3 D_0(\omega)$ and $C=15 e^{_{1}\over^{\sqrt 8 }}$ and
$c=\frac{1}{3}(\frac{2}{\sqrt 8}-\frac{1}{4})$.\hfill$\Box$

\vs 2mm

Let $A_k$ be, for $k>0$,
 the interval $[2^kR,2^{k+1}R]$  and let $A_0=[0,2R]$.
Given an integer $n_0$ define, for $k, k' \geq 0$,
 \qq
P(k | k') = \sup_{\omega' (0)}
P(D_{{n}_0} (\omega))\in A_k | \omega' (0))
\equiv
\sup_{\omega' (0)}
P(k| \omega' (0))
\label{d1}
\qqq
where the supremum is taken over $\omega'(0)$ such that
$\|\omega' (0)\|^2 \leq 2^{k'+1} R$ (the intervals
labelled by $k$ will play a role similar to the $k$'s
introduced in (\ref{chik}), but, since we do not need a
smooth partition of unity here, we use a more
conventional partition). Observe that we have $\forall k,
k'
\geq 0$,
\qq
 P (k | k') \leq 1 .
\label{d31}
\qqq
The main ingredient in the  proof of Proposition 2 is

\vs 2mm

\no{\bf Lemma 3.3.} {\it
There exist constants $ c > 0$, $C < \infty$ such that
\qq
P(k|k') \leq C \exp (-c2^k) \exp (e^{-(n_0-1)} 2^{k'-1})
\label{d41}
\qqq
}
\vs 3mm

\noindent
{\bf Proof.}  We split
\qq
&&P(k | \omega' (0))
=E\Bigl(1_{A_k} (D_{{n}_0} (\omega))
1(\|\omega (n_0-1)\|^2 > {_{2}\over^{3}} 2^k
R ) | \omega' (0) \Bigr)\non\\&&+ E\Bigl(1_{A_k} (D_{{n}_0} (\omega))
1(\|\omega (n_0-1)\|^2 \leq {_{2}\over^{3}} 2^k
R ) | \omega' (0) \Bigr),
\non
\qqq
where $1_{A_k}$ is the indicator function of the interval
$A_k$, and $1(X)$ is the indicator function of the event
$X$. Hence,
 we may bound
\qq
P(k|k') \leq \sup P \Bigl(\|\omega (n_0-1)\|^2 > {_{2}\over^{3}} 2^k
R | \omega'(0)\Bigr) + \sup E \Bigl(1_{A_k} (D_{{n }_0}
(\omega)) | \omega (n_0-1)\Bigr),
\label{d5}
\qqq
where the supremum in the first term is taken over $\omega'(0)$
such that $\|\omega' (0)\|^2 \leq 2^{k'+1} R$  and, in the
second term, over $\omega(n_0-1)$ such that $\|\omega (n_0-1) \|^2
\leq {2\over 3} 2^k R$.

Using Lemma 3.1, we bound the first term of (\ref{d5}) :
\qq
P\Bigl(\|\omega (n_0-1)\|^2 > {2\over 3} 2^k R |
\omega'(0)\Bigr) \leq 3 \exp (- {2^k\over 6}) \exp (e^{-(n_0-1)}
2^{k'-1}).
\label{d6}
\qqq
And, using Lemma 3.2, and the fact that the support of
$1_{A_k}$ is in $[2^k R, 2^{k+1}R]$ for $k>0$, we bound
the second term of (\ref{d5}), for $k>0$,   by
\qq
E \left(1_{A_k} \Bigl(D_{{n }_0} (\omega)\right) |
\omega (n_0-1)\biggr)\leq P
\Bigl(\sup_{t\in[n_0-1,n_0]} D_t (\omega) \geq 2^k R |
\omega (n_0-1)\Bigr)\leq C \exp (-c2^k),
\label{d7}
\qqq
since $\omega (n_0-1)$ is such that
$2^k R \geq \frac{3}{2}\|\omega (n_0-1)\|^2=3D_0(\omega) $.
For $k=0$, (\ref{d7}) obviously holds also. This proves
(\ref{d41}).\hfill$\Box$

\vs 3mm

\no {\bf Proof of  Proposition 2}. By Lemma 3.3, we may find
$n_0$ so that  $\exists \ c > 0$, $C < \infty$ such that
\qq
P(k|k') \leq C \exp (-c2^k) \ \ \mbox{for} \ k \geq k'.
\label{d4}
\qqq
Let us fix such $n_0$. Let $\CD$ be the sum of $D_n$ in
(\ref{x1}) and $\CD_\tau$ the same sum with $n$
restricted to the lattice $n_0\NZ+\tau$. We can write:
$$
P \Bigl(\CD
\geq \beta R|t'-t| \Big|
\omega(0) \Bigr)\leq \sum_{\tau=0}^{n_0-1}
P \Bigl(\CD_\tau
\geq \frac{\beta R|t'-t|}{n_0}
\Big|
\omega(0) \Bigr).
$$
So, since $|t'-t| \geq 1$, by
changing the values of $c$, and
$\beta_0$ in (\ref{x1}), it suffices to prove
(\ref{x1}) for $\CD$ replaced by $\CD_\tau$, $\tau=0,\dots,n_0-1$;
and, since all the terms are similar,
we shall consider only $\tau=0$.
Finally, by redefining $t$, $t'$, it is enough
to bound by the RHS of (\ref{x1})
the probability of  the event
\qq
\sum^{t'-1}_{n=t} D_{ nn_0}
(\omega) \geq \beta R|t'-t|.
\non
\qqq
Using the
Markov property, the definition (\ref{d1}) of $P(k|k')$,
and the fact that
$D_{nn_0}\in A_k$
means that
$ D_{nn_0}\leq 2^{k_{nn_0}+1}R$,
we see that it suffices (changing again $c$ and  $\beta_0$)
to prove the estimate
(\ref{x1}) for the expression
\qq
\sum_{\{k_{nn_0}\}} 1 \Bigl(\sum^{t'-1}_{n=t} 2^{k_{nn_0}} \geq \beta
|t-t'|\Bigr) \prod^{t'-2}_{n=t} P (k_{(n+1)n_0} | k_{nn_0})
P(k_{tn_0}| \omega (0)).
\label{d2}
\qqq
We bound (\ref{d2}), using Chebychev's
inequality, by
\qq
(\ref{d2})
\leq \exp(-\varepsilon \beta
|t'-t|)\sum_{\{k_{nn_0}\}} \exp (\varepsilon \sum^{t'-1}_{n=t} 2^{k_{nn_0}})
 \prod^{t'-2}_{n=t} P (k_{(n+1)n_0} | k_{nn_0})P(k_{tn_0}| \omega (0))
\label{d07}
\qqq
where $\varepsilon$ will be chosen small below.

Consider now $\sum_{k}\exp (\varepsilon  2^{k})
P (k | k')$.
Splitting this sum into $\sum_{0\leq k\leq k'-1 }$ and $\sum_{k\geq k'
}$
and using (\ref{d31}) for the first sum and (\ref{d4}) for the second,
we get:
\qq
\sum_{k}\exp (\varepsilon  2^{k})
P (k | k')\leq  k'\exp (\varepsilon  2^{k'-1}) +e^a
\label{d8}
\qqq
where $e^a \equiv C\sum_{k=0}^\infty \exp((\varepsilon -c) 2^{k})$
is bounded as long as (say) $\varepsilon\leq c/2$.
Moreover, we can bound $k'\exp (\varepsilon  2^{k'-1})+e^a
\leq e^{c_1} \exp (\varepsilon 2^{-\hf}
2^{k'})$. Altogether, we have:
\qq
\sum_{k}\exp (\varepsilon  2^{k})
P (k | k')\leq e^{c_1} \exp (\varepsilon 2^{-\hf}
2^{k'})
\label{d81}
\qqq
Let us apply this first to the sum over $k_{(t'-1)n_0}$, then $k_{(t'-2)n_0}$ and
so on.
The result of (\ref{d81}) is that, apart from the prefactor $e^{c_1}$,
we obtain, when we
sum over
$k_{(t'-2)n_0}$,
the same summand as in the first sum,
but with $\varepsilon$ replaced by $\varepsilon +
\varepsilon 2^{-\hf}$.  And, after $m$ steps we have
$\varepsilon$ replaced by $\varepsilon \sum_{l=0}^m 2^{-\hf l}$
Thus, we can use this  inductively
on $P (k_{(n+1)n_0} | k_{nn_0})$ for all $n$, with $t\leq n\leq t'-2 $, as long as
$\varepsilon \sum_{l=0}^\infty 2^{-\hf l}= \varepsilon (\frac{1}{1-
2^{-\hf}})\leq c/2$,
which holds for $\varepsilon$ small enough. Thus, we obtain, $\forall
t'>t$,
a bound for the sum in (\ref{d07})
\qq
e^{c_1|t'-1-t|}
\sum_{k_{tn_0}} \exp (c_2  \varepsilon 2^{k_{tn_0}})
  P(k_{tn_0} | \omega (0))
\label{d9}
\qqq
with $c_2 =\frac{1}{1- 2^{-\hf}}$.
Observe that, using  (\ref{d1}) and (\ref{d41}), with ${n}_0$ replaced
by ${t}$ and
 $k'$ being the smallest $k$ such that
$\|\omega (0)\|^2 \leq 2^{k+1} R$,
  we may bound
$$ P(k_{tn_0}| \omega
(0)) \leq C\exp (-c2^{k_{tn_0}} ) \exp (e e^{-t} 2^{k'-1}).
$$
Then the sum over $k_{tn_0}$ in (\ref{d9}) can be bounded, since
$\sum_{k_{tn_0}} \exp ((c_2  \varepsilon- c) 2^{k_{tn_0}})\leq C$ for
$\varepsilon$
small, and we get:
\qq
\sum_{k_{tn_0}} \exp (c_2 \varepsilon  2^{k_{tn_0}})
  P(k_{tn_0}| \omega (0))
\leq C\exp({e} e^{-t} 2^{k'-1} ).
\non
\qqq
Moreover, we have, by definition of $k'$, $2^{k'} \leq c\frac{\|\omega
(0)\|^2}{R}$.
Thus, we obtain the bound
 (\ref{x1}) for (\ref{d2}), for $\beta_0$ large enough
(e.g. take $\hf \varepsilon \beta \geq \hf \varepsilon \beta_0
\geq c_1 +\log C$, use $|t'-t|\geq 1$, and, in  (\ref{x1}),
take $c=\frac{\varepsilon}{2}$), by combining these
inequalities with
 (\ref{d07}) and  (\ref{d9}).\hfill$\Box$

\section{Partition of the path space}

Consider the expression (\ref{PT}) for the measure $\mu$.
 Given $\Nk$, we will now decompose the time axis into
regions where the equation (\ref{leq}) may have sensitive dependence on
initial conditions and the complement of those regions.
Motivated by Proposition 2, let
us consider, for time intervals $L$, the expressions
\qq
\gamma_L=\sum_{\Nn\subset L}
2^{k_n} .
\label{gammaL}
\qqq
Let $T$ be a number to be fixed later (in Sections 6-8), depending on $\rho$, the
minimum of the noise covariance.  Define
\qq
\beta (L)= \left\{\begin{array}{ll}
\beta |L|&\mbox{if $|L|>\hf T$}\\
\hf\beta T&\mbox{if $|L|\leq\hf T$}
\end{array}\right.
\label{larg}
\qqq
$\beta$ is
a constant to be fixed later (see Remark 2 below).
Call the time intervals
with end points on the lattice
$T\NZ$ $T$-intervals, and, for an interval $L=[m,n]$, let $\bar L$
be the smallest $T$-interval containing $[m,n]$ .
Consider
the set $\CL$ of intervals $L$ such that, either
\qq
\gamma_{L}>\beta (L),
\label{large}
\qqq
or $L=[(n-1)T,nT]$, so that
\qq
2^{k_{nT}}>\beta'T,
\label{large'}
\qqq
where $\beta'<\beta$ is a constant also to be fixed later
(see Remark 2 below). Let $\bar \CL$ be the union of all
$\bar L$ with $L\in\CL$. We call the connected components
of $\bar \CL$ {\it large} intervals and the $T$-intervals
of length $T$ in its complement {\it small} intervals.
Note that intervals of length $T$ can be either small or
large (those of length at least $2T$ are always large).
Hence, we introduce {\it labels } small/large on those
intervals. By construction, two large intervals are
always separated by at least one small one.

\vs 2mm

\no {\bf Remark 1.} ``Large'' and ``small'' refer to $\omega(J)$
being large or small, not to the size of the interval. We use
this slightly misleading terminology for the sake of brevity.
$\gamma_L$ are the natural random variables entering
in the sensitive dependence estimate (\ref{21}) and whose
probability distribution was studied in Proposition 2.
Since  the estimate (\ref{x1}) involves
 the initial condition at the beginning of the
time interval we consider and, since this initial
condition is the size of $\omega$ at the end of a time
interval where (\ref{large}) is violated, we need to be
sure it does not dominate the bound (\ref{x1}). For that
reason, we include in our set of unlikely events also the
ones defined by (\ref{large'}).

\vs 2mm

\no {\bf Remark 2.} The three constants in our construction,
$\kappa, \beta, \beta'$ entering (\ref{kappa}), (\ref{large})
and (\ref{large'}) are fixed as follows: $\beta'\geq\beta'_0$,
$\beta\geq\beta(\beta')$ and $\kappa\geq \kappa(\beta)$.

\vs 2mm

\no {\bf Remark 3.} The virtues of this partition of phase space
can be seen in Lemma 4.1 and 7.4 below. The bound
(\ref{large2}) and Proposition 2 will imply that large
intervals are unprobable. On the other hand,
(\ref{small}) and (\ref{large11}) will allow us to show
that the argument of the exponential in (\ref{21}) is
less than $-cRT$, when the interval $[0, t]$ is replaced
by an interval strictly including one of the intervals
constructed here. This property will be essential in
order to obtain bounds on the terms of the expansion
constructed in the next Section.

\vs 2mm

Taken together, the small and large
intervals form a  partition $\pi(\Nk)=J_1,\dots ,J_N$
of the total time interval $[0,t]$. We arrange them in temporal
order and write $J_i=[\tau_{i-1},\tau_i]$ with
$\tau_0=0$,  $\tau_N=t$.

Our construction has the following
properties

\vspace*{2mm}

\no{\bf Lemma 4.1.} {\it  Let $J=[\tau',\tau]$ be a $T$-interval
$J\in \pi(\Nk)$.  }

\no (a) {\it If $J$ is small, then}
\qq
\sum_{\Nn\subset J}2^{k_n}\leq \beta T \; {\rm and} \;
2^{k_{\tau}}\leq \beta'T,
\label{small}
\qqq

\no (b) {\it If $J$ is large, then $J$ may be written as
a union $J'\cup J''$ so that
\qq
\gamma_{J'} > {_{1}\over^{4}}{\beta} |J'|
\label{large2}
\qqq
and $J''$ is a union of
intervals $[(n-1)T,nT]$ satisfying (\ref{large'}).}

\vs 2mm

\no {\bf Remark 4.} At both ends of any interval, either large or small,
we have $2^{k_n}\leq \beta T$ (otherwise the interval would be large, not small,
or would not end there). Note that we have $\beta$ here, not the smaller $\beta'$
of (\ref{small}). So, if $\omega$ is such that $D_n(\omega)$ is in the support
of $\phi_{k_n}$, we have:
\qq
\|\omega(\tau)\|^2\leq 8\beta R T
\label{s60}
\qqq
where $\tau$ is the endpoint of the interval.
\vs 2mm

\no{\bf Proof of Lemma 4.1}.
(a) A small interval cannot be an
$L$ for which (\ref{large}) holds nor  an interval $[(n-1)T,nT]$
 satisfying (\ref{large'}); hence, (\ref{small}) holds.

For (b), let $J'$ be the union of the $\bar L$
in  $J$ with $L$ such that  (\ref{large}) holds. We may cover
$J'$ by a subset $\bar L_{i}$, $i=1,\dots p$, of these intervals,
in such a way
that $\bar L_{i}\cap\bar L_{j}=\emptyset$ for $|i-j|>1$. From
(\ref{large}, \ref{larg}),
we deduce that $\gamma_{\bar L}>{_{1}\over^{2}}\beta|\bar L|$ and then,
$$
\gamma_{J'}\geq \hf\sum\gamma_{\bar L_{j}}\geq
{_{1}\over^{4}}\beta|J'|.
$$
\hfill $\Box$

\vs 2mm

In order to obtain the analogue of what in Statistical
Mechanics is called the high temperature expansion, we
need to write the sum in (\ref{PT}) as a sum of products
of independent factors. As a first step in that
direction, we would like to express the sum in (\ref{PT})
as a sum of partitions $\pi=(J_1,\dots,J_n)$ of $[0,t]$
into $T$-intervals and sums over $\Nk_i\in\NN^{J_i}$.
However, a moment's thought reveals that the sum over
$\Nk$ creates correlations between the different $\Nk_i$.
E.g. $J_i$ being small is a very nonlocal condition in
terms of $\Nk$: nowhere in the whole interval $[0,t]$ can
there be a $k_n$ large enough to create a $L\in \CL$ that
intersects $J_i$.  Given an arbitrary $T$-interval $J$ and
$\Nk\in\NN^J$, we may define, in the same way as we did
above for $[0,t]$, the partition $\pi(\Nk)$ of $J$ into
small and large intervals. In particular,
$\pi(\Nk)=\{J\}$ means, if $|J|=T$, that $\Nk$ is such
that $J$ is small or large depending on the label on $J$
and, if $|J|>T$, that $\Nk$ is such that $J$ is large.
Then, we have:

\vspace*{3mm}

\no{\bf Lemma 4.2.} {\it
Let $\pi=\{J_1,\dots ,J_N\}$ be a partition of $[0,t]$
into $T$-intervals and let $\Nk_i\in \NN^{J_i}$ be given such  that
$\pi(\Nk_i)=\{J_i\}$.  Let $\Nk=(\Nk_1,\dots, \Nk_N)$.
Then $\pi(\Nk)=\cup_i\pi(\Nk_i)$ if and only if the $\Nk_i$
satisfy the constraints
\qq
\forall L\subset J_i\cup J_{i+1}\;{\rm so \; that}\;
L\cap J_i\neq\emptyset\neq L\cap J_{i+1}\; :\;\; \gamma_L\leq\beta(L)
\label{con1}
\qqq
for all $i=1,\dots, N-1$}

\vspace*{3mm}

\no{\bf Proof.}
Assume first that $\pi(\Nk)=\cup_i\pi(\Nk_i)$. Hence
$\pi(\Nk)=\pi$ and by the definition of $\pi(\Nk)$, every
$L$ such that $\gamma_L>\beta(L)$ is contained in some
$J_i$. Thus, (\ref{con1}) holds.

For the converse, observe first that, by the definition
of the partitions $\pi(\Nk)$ and $\cup_i\pi(\Nk_i)$,
their sets of small and large $J$'s are entirely
determined by the set of connected components  of ${\cal
L}$ given by $\Nk$ on $[0,t]$ for $\pi(\Nk)$ and the set
of connected components of ${\cal L}_i$ given by $\Nk_i$
on each $J_i$ for $\cup_i\pi(\Nk_i)$. Thus it is enough
to show that their connected components coincide.  The
intervals satisfying (\ref{large'}) obviously coincide.
By definition of $\gamma_L$ and of the large intervals,
each connected component of ${\cal L}_i$ must be
contained in a connected component of ${\cal L}$, since
$\Nk=(\Nk_1,\dots, \Nk_N)$. Now, using  (\ref{con1}), we
show the converse, which will establish the claim.  Let
$L$ be a connected component of $\CL$. If $L\subset J_i$,
then $L$ is a connected component of $\CL_i$. Thus, if
the claim is not true, there must exist a connected
component $L$ of $\CL$, such that  $L$ is not included in
any $J_i$ and such that $\gamma_L>\beta(L)$.  By
(\ref{con1}), $L$ cannot be included in two adjacent
$J$'s either. Thus, there must be a connected $L$ with
$\gamma_L>\beta(L)=\beta |L|$ and $J_i$ such that $L\cap
J_{i-1}
\neq \emptyset $, $L\cap J_{i+1} \neq \emptyset $.  Then
$L=L_1\cup L_2$ with $L_1, L_2$ having the midpoint of
$J_i$ as a common boundary point. Hence, by (\ref{con1}),
$\gamma_{L_i}\leq
\beta(L_i)=\beta|L_i|$ since $|L_i|>\hf T$. Thus
$\gamma_L=\gamma_{L_1}+\gamma_{L_2}\leq \beta |L|=\beta(L)$, which is a
contradiction. \hfill$\Box$

\vspace*{3mm}

Consider now the sum (\ref{PT}). Let $\pi(\Nk)=\{J_1,\dots,J_N\}$.
Define the Girsanov factor
\qq
g_{J_i}(\omega)=e^{\int_{J_i}(f(t),\gamma^{-1}(ds(t)-\hf f(t)dt))}
\label{girs1}
\qqq
where we recall that $f(t)$ and $\omega$, given by
(\ref{f}) and (\ref{oomega}), and thus $g_{J_i}$, depend
on the whole past i.e. on ${\bf s}([0,\tau_i])$
and $l(0)$.  Let $\Nk_i\in \NN^{J_i}$ be the restriction of $\Nk$ to
$J_i$,  let us denote by $\chi_{\Nk_i}$ the corresponding product
(\ref{chik}), and let
\qq
\mu_{\Nk_i}(d\Ns(J_i))=\chi_{\Nk_i}
g_{J_i}\nu_{s(\tau_{i-1})}^{|J_i|}(d\Ns(J_i)).
\label{mui}
\qqq
We can then write
\qq
\chi_{\bf k}\mu^t_{\omega(0)}(d\Ns)=
\prod_{i=1}^N\mu_{\Nk_i}(d\Ns(J_i)).
\label{coupled}
\qqq
Let $\pi$ be a partition of $[0,t]$ into $T$-intervals
with labels ``small'' or ``large'' on the ones of
length $T$.
Let us define, for such a labelled  $T$-interval $J$,
$1_{J}(\Nk)$ to be
the indicator function for the set of $\Nk\in\NN^J$
such that $\pi(\Nk)=\{J\}$ (i.e. if $|J|=T$ $1_J$ is supported on
$\Nk$ so that $J$ is small or large depending on the label and
if $|J|>T$  on
$\Nk$ so that $J$ is large). For two adjacent $T$-intervals
$J,J'$ let $1_{JJ'}(\Nk,\Nk')$  be
the indicator function for the set of $(\Nk,\Nk')\in
\NN^J\times \NN^{J'}$,
such that $\gamma_L\leq\beta(L)$ for all $L\subset J\cup J'$
which intersect both $J$ and $J'$.
Using Lemma 4.2, we may then write  eq. (\ref{PT})
as
\qq
\mu^t_{\omega(0)}(d\Ns)=\sum_\pi\sum_{\Nk_1\dots\Nk_N}
\prod_{i=1}^N 1_{J_i}(\Nk_i)\mu_{\Nk_i}(d\Ns(J_i))
\prod_{i=1}^{N-1}1_{J_iJ_{i+1}}(\Nk_i,\Nk_{i+1}).
\label{coupled1}
\qqq
Note that this expression has a Markovian structure
in the sets $J_i$, but each $\mu_{\Nk_i}$
depends on the whole past history. In the next Section,
we shall decouple this dependence.

\section{Decoupling}

By decoupling we mean that we shall write
$\mu^t_{\omega(0)}$ as a  product of measures whose
dependence on the past extends only over two adjacent
intervals,  and corrections. To achieve that, consider
$\mu_{\Nk_i}$, for $i>2$; remember that $[0,t]$ is
partitioned into intervals $J_i=[\tau_{i-1},\tau_i]$ with
$\tau_0=0$, $\tau_N=t$. Fix $j<i$, and introduce the
drift with memory on $[\tau_{j-1},t]$ :
\qq
f_j(t)=PF(\omega_j(t))
\label{ftau}
\qqq
where
$$
\omega_j(t)=(s(t),l(t,{\bf s}([\tau_{j-1},t]),0))
$$
is the solution of (\ref{seq}, \ref{leq}), with initial
condition $l(\tau_{j-1})=0$. We denote by $g_{ij}$ the
Girsanov factor $g_{J_i}(\omega_j)$ (given by
(\ref{girs1}), with $f(t)$ replaced by $f_j(t)$). Note
that it depends only on the history ${\bf
s}([\tau_{j-1},\tau_{i}])$.

Since the
characteristic function $\chi_{\Nk_i}$ also depends on the past through
the $\omega$ dependence of  (\ref{chik}), we need to decouple this too.
We let
\qq
\chi_{\Nk_ij}=\prod_{\Nn\subset J_i}\phi_{k_n}(D_n(\omega_j)).
\label{chik1}
\qqq

We can now define the decoupled measure for
$j=2,\dots,i-1$:
\qq
\mu_{\Nk_ij}(d{\bf s}(J_i)|{\bf s}([\tau_{j-1},\tau_{i-1}]))=\chi_{\Nk_ij}
g_{ij}\nu_{s(\tau_{i-1})}^{|J_i|}(d\Ns(J_i));
\label{muij}
\qqq
this measure is defined on the paths  on the time
interval $J_i$ and depends on the past up to and
including the interval $J_j$. To connect to
(\ref{coupled}), we write, for $i\geq 3$, a telescopic
sum
\qq
\mu_{\Nk_i}
=\mu_{\Nk_i i-1}+
\sum_{j=1}^{i-2}(\mu_{\Nk_ij}-\mu_{\Nk_ij+1})\equiv
\sum_{j=1}^{i-1}\mu_{\Nk_i,j}\;,
\label{mui,j}
\qqq
where by definition $\mu_{\Nk_i1}=\mu_{\Nk_i}$; note that
this term  is the only one depending  on $l(0)$. For
$i=1,2$ we will set by convention $j_i=i-1$,
$\Ns([\tau_{-1}, \tau_0]) =\omega (0)$, and
define
$\mu_{\Nk_ij_i}=\mu_{\Nk_i}$. Inserting (\ref{mui,j})
into (\ref{coupled1}), we get
\qq
\mu^t_{\omega(0)}(d\Ns)=\sum_\pi\sum_{\Nk_1\dots\Nk_N}\sum_{\Nj}
\prod_{i=1}^N 1_{J_i}(\Nk_i)\mu_{\Nk_i,j_i }(d{\bf s}(J_i)|{\bf
s}([\tau_{j_i-1},\tau_{i-1}]))
\prod_{i=1}^{N-1}1_{J_iJ_{i+1}}(\Nk_i,\Nk_{i+1}).
\label{z}
\qqq
One should realize that the leading term in the sum
(\ref{z}) is the one with all $j_i=i-1$ and $\Nk$ such
that the partition $\pi(\Nk)$ consists of only small
intervals. Indeed, $\mu_{\Nk_i,j_i }$ with $j_i\neq i-1$
describes the change of $\mu_{\Nk_i}$ under variation in
distant past. This will be shown to be small as a
consequence of Proposition 1 and Lemma 4.1. On the other
hand, the occurrence of large intervals
 will be shown to have a small probability,
using Proposition 2.

Therefore, we will group all these small terms as
follows. Consider the set
\qq
L'=\bigcup_{j_i<i-1}[\tau_{j_i-1},\tau_{i}]
\bigcup_{J_i\; {\rm large}}(J_i\cup J_{i+1})
\label{K'alpha}
\qqq
where we have grouped the terms mentioned above, and also
included the small intervals following the large ones for
later convenience. Since our initial condition
$\omega(0)$ is arbitrary, it is  convenient to include
also the intervals $J_1, J_2$, and to let
\qq
L=J_1\cup J_2\cup L'.
\label{Kalpha}
\qqq
Let $K_1,\dots ,K_N$
be the partition of $[0,t]$ into $T$-intervals,
in chronological order, where the $K_l$'s, which are
unions of intervals $J_i$,
are given by the connected components of $L$ and by the
small intervals $J_i\subset L^c$. In
the first case, $|K_l|\geq 2T$,
since we always attach
to a large interval $J_i$
the interval $J_{i+1}$,
 see (\ref{Kalpha}, \ref{K'alpha});
 in the second case,
$|K_l|=T$.

Fix now $K=[\tau_0,\tau]$,  a $T$-interval and let $J_0=
[\tau_0-T,\tau_0]$ if $0\notin K$. Let $\Nk_+, \Nk_0\in
\NN^T$ and $\Ns(J_0) \in C (J_0, H_s)$. We define
\qq
&&\mu_{K}(d\Ns(K), \Nk_+|\Ns(J_0), \Nk_0)=\non\\
&&\sum_\pi\sum_{\Nk_1\dots\Nk_{N-1}}\sum_{\Nj}
\prod_{i=1}^N 1_{J_i}(\Nk_i)\mu_{\Nk_i,j_i }
(d{\bf s}(J_i)|{\bf
s}([\tau_{j_i-1},\tau_{i-1}]))
\prod_{i=0}^{N-1}1_{J_iJ_{i+1}}(\Nk_i,\Nk_{i+1})
\label{muKk}
\qqq
where $\Nk_N=\Nk_+$, for $i=1$, ${\bf
s}([\tau_{j_i-1},\tau_{i-1}]$ is replaced by $\Ns (J_0)$,
and the sum is over $\pi$ and
$\Nj$ so that $K$ equals $L'$ of (\ref{K'alpha}) if $0\notin
K$, or $L$ of (\ref{Kalpha}) if $0\in K$. In the latter
case, we replace $\Ns(J_0), \Nk_0$ by $\omega(0)$ and the
last product starts at $i=1$. Note that, because
of the presence of $J_{i+1}$ in (\ref{K'alpha}),
the last interval in $K$ is small.
With this definition, we
can then rewrite (\ref{z}) as:
\qq
\mu^t_{\omega(0)}(d\Ns)=\sum_\pi
\sum_{\Nk_1\dots\Nk_{M}}
\prod_{i=1}^M\mu_{K_i}
(d\Ns(K_i), \Nk_i|\Ns(J_{i-1}), \Nk_{i-1})
\label{z1}
\qqq
where the sum is over partitions $\pi=(K_1,\dots, K_M)$
of $[0,t]$ into $T$-intervals $K_i=[\tau_{i-1},\tau_i]$
so that $|K_1|\geq 2T$ (because we included $J_1,J_2$
into $K_1$, see (\ref{Kalpha})), and for $i=1$,
$\Ns(J_{i-1}), \Nk_{i-1}$ is replaced by $\omega_0$. Note
that all the $K_l$'s so that $|K_l|=T$
 are
small intervals, and, in that case,
$K_l$  coincides with an
interval $J_i=[\tau_{i}-T,\tau_i]$.

\vs 2mm

The expansion in (\ref{z1}) has a  Markovian structure in
the pairs $\sigma=(\Ns(K),\Nk)\equiv (\Ns,\Nk)$, and it
is convenient to set
\qq
\mu_{K}
(d\sigma|\sigma')=\mu_{K} (d\Ns, \Nk|\Ns', \Nk').
\label{musigma}
\qqq
We write for the convolution of such kernels:
\qq
\mu_{K}\mu_{K''}(d\sigma|\sigma'')=
\int\mu_{K}(d\sigma|\sigma')\mu_{K''}(d\sigma'|\sigma'')
\label{convo}
\qqq
where the integral means both the integral over $d\Ns'$
and the sum over $\Nk'$. For $|K|=T$, i.e. for a small
interval, we drop the index $K$ altogether in our
notation and write $\mu^n$ for the $n$-fold convolution.
With these preparations, let us then consider the
expression (\ref{girs}) when the function $F$ depends
only on $s$:
\qq
P^t(\omega(0),F)=\int \mu^t_{\omega(0)}(d\Ns)F(s(t))=
\sum_{\NK,\Nn}\int \mu^{n_M}
\mu_{K_{M}}\dots \mu^{n_1}
\mu_{K_{1}}(d\sigma|\omega(0))F(s(t))
\label{PT1}
\qqq
where  $\NK=(K_1,\dots,K_M)$ are disjoint
$T$-intervals of length at least $2T$, $\sum |K_i|+T\sum n_i=t$,
$n_M\geq 0$, $n_i>0$ for other $i$'s and $M\geq 1$.

There are two kinds of transition kernels in (\ref{PT1}),
the unlikely ones $\mu_K$ and the likely ones $\mu^n$.
The latter will be responsible for the convergence to
stationarity and we will discuss them next. Let
$\sigma=(\Ns,\Nk)$ with $\Ns=\Ns(J)$ and
$J=[\tau,\tau+T]$, $J_0=[\tau-T,\tau]$. Define
\qq
P(d\Ns|\Ns')=g_J(\omega)\nu^T_{s(\tau)}(d\Ns)
\label{Pss'}
\qqq
where $\omega(t)=(s(t), l(t,\Ns\vee\Ns'([\tau-T,t]),0)$,
with $\Ns\vee\Ns'$ being the configuration on
$[\tau-T,\tau+T]$ coinciding with $\Ns'$ on
$[\tau-T,\tau]$, and with $\Ns$ on $[\tau,\tau+T]$; we
put $P(d\Ns|\Ns')=0$ if $s(\tau) \neq s'(\tau)$;
$g_J(\omega)$ is the Girsanov factor (\ref{girs1}) (which
here, of course,
 because of the
definition of $\omega$, depends only on $\Ns\vee\Ns'([\tau-T,t])$).
Let also
\qq
\chi_\Nk(\Ns,\Ns')=\chi_\Nk(\omega)
1_J(\Nk),
\label{chiss'}
\qqq
where $1_J(\Nk)$ is supported on $\Nk$ so that $J$ is a
small interval. Then, (\ref{musigma}) in the special case
$|K|=T$ gives:
\qq
\mu(d\sigma|\sigma')=\chi_\Nk(\Ns,\Ns')
P(d\Ns|\Ns')1_{J_0J}(\Nk',\Nk).
\label{muk1}
\qqq
 Let $\bar \mu$ be
given by (\ref{muk1}) without the $1_{J_0J}(\Nk',\Nk)$
factor:
\qq
\bar\mu(d\sigma|\sigma')=\chi_\Nk(\Ns,\Ns')P(d\Ns|\Ns')
\label{barmu}
\qqq
 and write
\qq
\mu=\bar\mu+\Delta.
\label{Delta}
\qqq
$\Delta(d\sigma|\sigma')$ is a  measure of small total mass,
since it is supported on $\sigma$'s such that
large intervals $L$ intersect two adjacent small ones.
So, let us expand:
\qq
\mu^n=(\bar\mu+\Delta)^n=\sum\bar\mu^{n_1}\Delta^{n_2}\dots
\bar\mu^{n_{k-1}}\Delta^{n_k}.
\label{mu^n}
\qqq
We will state now the basic bounds for the transition
kernels that allow us to control the expansions
(\ref{muk1}) and (\ref{mu^n}). Remember that the initial
states $\Ns'$ in our kernels are on small intervals
$J_0=[\tau-T,\tau]$ (except for the $\mu_K$ with $0\in K$
which has $\omega(0)$ as initial state). This means that
$\omega'(t)=\omega(t,{\bf s}'([\tau-T,t]),l'(\tau-T))$ is
constrained to be on the support of the $\chi_\Nk$ with
$\Nk$ such that $J_0$ is small.
This implies  that all the transition kernels have
initial states
 $\Ns' \in C_s\subset C(J_0 ,H_s)$
given by (see (\ref{chik}) and the support of $\phi_k$)
\qq
C_s=\{\Ns'\;|\;\sum_{\Nn\subset J_0}D_n(\omega')\leq
4\beta RT\; ,\; D_{\tau}(\omega')\leq 4\beta' RT\}.
\label{Cs}
\qqq
\vs 2mm

\vs 2mm

The first Proposition controls the unlikely events of having
either $\Delta^n$, $n\geq 1$, or $\mu_K$ with $|K|\geq 2T$ (or both):

\vspace*{3mm}

\no{\bf Proposition 3.}  {\it
There exists $c>0$, $c'<\infty$, $T_0=T_0(\rho,
R)<\infty$ such that, $\forall T\geq T_0$, and for
$|K|\geq 2T$, or $m\geq 2$, or $m=1$ and $|K|\geq T$,
\qq
\sup_{\Nk'}\sup_{\Ns'\in C_s}\int|\Delta^m\mu_{K}(d\sigma|\sigma')|\leq
e^{-c(|K|+Tm)}
C_K(\omega(0))
\label{muk}
\qqq
where the sup is over $\Nk'$ so that $J_0$ is small, if
$0\notin K$. $C_K(\omega(0))=1$ if $0\notin K$ and
\qq
C_K(\omega(0))= e^{c'\beta' T}e^{
\frac{||\omega(0)||^2}{8R}}
\label{CK}
\qqq
if $0\in K$.
}

\vspace*{3mm}

For the likely events we look more closely at $\bar\mu^n$:
\qq
\bar\mu^n(d\sigma|\sigma')=\int\bar\mu(d\sigma|\sigma'')
\lambda^{n-1}(d\Ns''|\Ns')
\label{barmun}
\qqq
with $\lambda$ given by, see (\ref{barmu}),
\qq
\lambda(d\Ns|\Ns')=\sum_\Nk\bar\mu(d\sigma|\sigma')
=\sum_\Nk\chi_\Nk(\Ns,\Ns')P(d\Ns|\Ns').
\label{muss'}
\qqq
The content of the following proposition is that
 $\lambda^n$ relaxes to equilibrium:

\vs 3mm

\no{\bf Proposition 4.}
{\it There exist $\delta=\delta(\rho, R)>0$,
$p=p(\rho, R)<\infty$,
such that, $\forall T\geq T_0$,
\qq
\sup_{\Ns'\in C_s}\int|\lambda^p(d\Ns|\Ns')-
\lambda^p(d\Ns|0)|\leq 1-\delta
\label{mun}
\qqq
}

\section{Proof of the Theorem}

The proof of the Theorem is rather straightforward, given
the estimates, stated in Propositions 3 and 4, on the
measures in (\ref{PT1}). Note that the length $T$ of the
intervals entering in the expansion (\ref{PT1}) is a
parameter that has not yet been fixed. For simplicity, we
shall consider only times $t$ in (\ref{con}) that are
multiples of $T$; the general case is easy to obtain.

We divide the proof into two parts: in the first one,
$F=1$ in  (\ref{con}) and, in the second, $F$ is a
general H\"older continuous function.

In the case $F=1$, we integrate a function, $1_E$,
depending only on $s'$ and we may use (\ref{PT1}). Let
$$
\mu_0(d\Ns)\equiv\lambda^p(d\Ns|0)
$$
and rewrite (\ref{mun}) as
\qq
\sup_{\Ns'\in C_s}\int
|\lambda^p(d\Ns|\Ns')-
\mu_0(d\Ns)|\leq 1-{ \delta}.
\label{mun1}
\qqq
In (\ref{PT1}), first, expand each
$\mu^{n_i}$ factor, for $i=1,
\ldots M$ using (\ref{mu^n}):
\qq
\mu^{n_i}=(\bar\mu+\Delta)^{n_i}=\sum\bar\mu^{n_{i1}}\Delta^{m_{i1}}\dots
\bar\mu^{n_{ik_i}}\Delta^{m_{ik_i}}.
\label{600}
\qqq
Then we write, using (\ref{barmun}),
\qq
{\bar \mu}^{ n_{ij}}={\bar \mu} (\lambda^p
-\mu_0 + \mu_0)^{[{{_{ n_{ij}-1}}\over^p}]}\lambda^{q_{ij}},
\label{601}
\qqq
 where $\lambda $ is defined by (\ref{muss'}) and
$ n_{ij}-1=[\frac { n_{ij}-1}{p}]p +q_{ij}$, i.e.
$q_{ij}<p$. Finally, expand each of the resulting factors
\qq
(\lambda^p
-\mu_0 + \mu_0)^{[{{_{ n_{ij}-1}}\over^p}]}=\sum M_{a_{ij}}
(\lambda^p-\mu_0,\mu_0)
\label{602}
\qqq
where $M_{a_{ij}}$ is a monomial, of degree $a_{ij}$ in
the first variable. This way we end up with an expansion
of $P^t(\omega(0),1_E)$ in terms of products of $\mu_K$
with $|K|\geq 2T$, $\Delta^m$, $\lambda^p-\mu_0 $, $\Delta\bar
\mu$, $\bar
\mu$, $\lambda^{q}$, with $q<p$,  and of $\mu_0$.

 Consider now two initial conditions
$\omega(0)=\omega_0,
\omega'_0$ and let  $||\omega'_0||\leq ||\omega_0||$.
Let $t_0 = {_{C}\over^{\beta R}}||\omega_0||^2+T$, and
perform the expansion (\ref{600}-\ref{602}) for the
factors $\mu^{n_i}$ that occur after $t_0$ in
(\ref{PT1}). Let, for $n\geq t_0$, $P^t_n(\omega_0,1_E)$
consist of all the terms in the resulting sum that have
$\mu_0(d\Ns)$ with $\Ns={\bf s}([(n-1)T,nT])$ as one of
the factors in the product. Note that, if $n$ is larger
than $t_0$,  such terms always  exist. Indeed, $D_1\geq
\hf ||\omega_0||^2$ forces $2^{k_1}\geq {_{1}\over^{8R}}
||\omega_0||^2$ and thus implies that the origin is
contained in a large interval of length
${_{C}\over^{\beta}}2^{k_1}$; but longer intervals are
not forced by the initial condition, and so, $\mu^{n_i}$
factors are not forbidden in  (\ref{PT1}), after $t_0$.
The same will be true for $\omega'_0$, since
$||\omega'_0||\leq ||\omega_0||$. Since, $\mu_0 (d\Ns) $
is independent of the past, the sum in
$P^t_n(\omega_0,1_E)$ factorizes and, for the times
before $(n-p)T$, we recover the full $P^{(n-p)T}$. We
have then
\qq
P^t_n(\omega_0,1_E)=\int
P^{(n-p)T}(\omega_0,d\Ns')\mu_0(d\Ns)
f(n,\Ns,E)=\int\mu_0(d\Ns) f(n,\Ns,E)
\non
\qqq
since $P^{(n-p)T}(\omega_0,H_s)=1$. Thus,
$$
P^t_n(\omega_0,1_E)=P^t_n(\omega'_0,1_E),
$$
and we conclude that
$$P^t(\omega_0,1_E)-P^t(\omega'_0,1_E)
=R^t(\omega_0,1_E)-R^t(\omega'_0,1_E)
$$
where $R^t(\omega_0,1_E)$ is given by the same sum as
$P^t(\omega_0,1_E)$ except for the terms that have a
factor $\mu_0(d\Ns)$ with $\Ns={\bf s}([(n-1)T,nT])$ for
$n\geq t_0$.

 We will estimate $|R^t(\omega_0,1_E)|$. Note that
it contains only, after time $t_0$, the factors $\mu_K$
with $|K|\geq 2T$, $\Delta^m$, $\lambda^p-\mu_0 $,
$\Delta\bar
\mu$, $\bar
\mu$, $\lambda^{q}$, with $q<p$, i.e. no $\mu_0$ factors.
Let us count the powers of the various factors in its
this expansion, using the definitions in eqs.
(\ref{600}), (\ref{601}) and (\ref{602}). The number
$N_\Delta$ of $\Delta$-factors is
\qq
N_\Delta=\sum_{ij} m_{ij}.
\non
\qqq
To count the number of $\lambda^p-\mu_0$ factors, note
that only the term with $a_{ij}=[{{_{
n_{ij}-1}}\over^p}]$ in (\ref{602}) enters (all the
others having at least one $\mu_0$);

\qq
&&\sum_{ij} a_{ij}= \sum_{ij}  {[{{_{
n_{ij}-1}}\over^p}]}
\geq {_{1}\over^{p}}\sum_{ij} (n_{ij}-2)
={_{1}\over^{p}}\sum_in_i-{_{1}\over^{p}}
\sum_{ij}(m_{ij}+2)
\label{604}
\qqq
where, in the last step, we used (\ref{600}). Since
$\sum_i 1 =M'$, where $M'$ is the number of
$K_i$ factors in (\ref{PT1}) that do not occur before $t_0$,
 we get
$\sum_{ij}(m_{ij}+2)\leq 3N_\Delta+2M'$, where the first
term bounds the sum over $m_{ij}\neq 0$, and the second
the sum over $m_{ij}=0$. Thus,
\qq
(\ref{604}) \geq {_{1}\over^{p}}\Bigl(\sum_in_i-3N_\Delta-2M'
\Bigr)
\geq {_{1}\over^{p}}\Bigl({_{t-t_0}\over^{T}}
-{_{1}\over^{T}}\sum_{i} (|K_i|+2)-3N_\Delta\Bigr)
\label{603}
\qqq
where in the last step we used $T\sum n_i+\sum
|K_i|\geq t-t_0$ (remembering that we use the expansion
in  (\ref{600}-\ref{602}) only after time
$t_0$).

In order to bound $|R^t(\omega_0,1_E)|$, which is a sum
of terms, we shall first bound all the factors in each
term. For $\mu_K$, for $|K|\geq 2T$, and $\Delta^m$,
$m\geq2$, we  use  (\ref{muk}) and, writing $\Delta {\bar
\mu}=
\Delta { \mu} -\Delta^2$, we obtain a bound like
(\ref{muk}) (with another $c$), for   $\Delta {\bar \mu}$
instead of $\Delta { \mu}$; for $\lambda^p-\mu_0$, , we
use (\ref{mun}). The  other terms have simple bounds:
since ${\bar
\mu}(d\sigma|\sigma')$, defined in (\ref{barmu}), is
positive, we have
\qq
\sup_{\sigma'}\int|{\bar \mu}
(d\sigma|\sigma')| =
\sup_{\sigma'}\int{\bar \mu} (d\sigma|\sigma')\leq1,
\label{x25}
\qqq
and, similarly, by (\ref{muss'}),
$$
\sup_{\Ns'}\int|\lambda^q
(d\Ns|\Ns')|\leq 1;
$$
We also have, for  $\Delta^m$ with $m=1$,
$$
\sup_{\sigma'}\int |\Delta (d\sigma|\sigma')| \leq
\sup_{\sigma'} \int|{\bar \mu}
(d\sigma|\sigma')|+ \sup_{\sigma'}\int|{\mu}
(d\sigma|\sigma')|\leq 2.
$$
Observe that the last three factors occur always next to
other factors: ${\bar \mu} $ or $\Delta$ at the beginning
or the end of the products in (\ref{600}) (actually,
there is, in the full expansion, at most one factor
$\Delta$ not multiplied by ${\bar \mu}$ or by $\mu_K$) or
$\lambda^q$ at the end of the product in (\ref{601}). So,
the summation in $R^t$ runs only over the sets $K_i$ in
(\ref{PT1}) and over the occurrences of $\Delta$ in
(\ref{600}) (since only the term without a $\mu_0$ factor
in (\ref{602}) enters in $R^t$). Combining this
observation, all  the above inequalities and (\ref{603}),
 we can bound $R^t$ by a sum of
\qq
 C(\omega(0))
(1-\delta)^{{_{1}\over^{p}}({_{t-t_0}\over^{T}})}
e^{-c(\sum |K_i|+TN_\Delta)},
\non
\qqq
over the subsets $K'$ consisting of the union of the
$T$-intervals $K_i$ and of the $T$-intervals where
$\Delta$ occurs; here, $C(\omega(0))=C_K(\omega(0))$,
given by (\ref{CK}).
Since $t_0$ depends on $\omega_0$, we may absorb the
factor $(1-\delta)^{{_{1}\over^{p}}({_{-t_0}\over^{T}})}$
into $C(\omega(0))$, and we get:
\qq
|R^t(\omega_0,1_E)|\leq  C(\omega(0))
(1-\delta)^{{_{1}\over^{p}}({_{t}\over^{T}})}
\sum_{K'}e^{-c|K'|}\leq C(\omega(0))
(1-\delta')^{{_{t}\over^{T}}}e^{e^{-cT} {_{t}\over^{T}}}
\label{x13}
\qqq
since the sums over subsets of $[0, t]$ made of
$T$-intervals  can be identified with sums over subsets
of $[0, t/T]$ (remember that $t$ a multiple of $T$).
$\delta'$ is defined by $1-\delta'=
(1-\delta)^{_{1}\over^{p}}$ and, like ${\delta}$ and $p$,
is independent of $T$. Therefore, choosing $T$ large
enough, (\ref{x13}) can be bounded by
$C(\omega(0))e^{-mt}$ for some $m>0$ depending on $R$ and
$T$, i.e. on $R$ and $\rho$ ($T$ will be chosen as a
function of $\rho$ in the next Section). Using a similar bound for
$R^t(\omega'_0,1_E)$, we obtain that
\qq
|P^t(\omega(0),1_E)-P^t(\omega'(0),1_E)|\leq
C(\omega(0))e^{-mt}
\label{x130}
\qqq
>From this, the existence of the limit $\lim_{t\to
\infty}P^t(\omega(0),1_E)$ follows: indeed, write, for $t>t'$,
\qq
P^t(\omega(0),1_E)-P^{t'}(\omega(0),1_E)
=\int P^{t-t'}( \omega(0), d\omega)
(P^{t'}(\omega, 1_E)-P^{t'}(\omega(0),1_E))
\label{x21}
\qqq
and use (\ref{x130})
$$
|P^{t'}(\omega,1_E)-P^{t'}(\omega(0),1_E)|\leq
(C(\omega)+C(\omega(0))) e^{-mt'}.
$$
Then we have, by (\ref{c5}) and (\ref{CK}),
\qq
\int P^{t-t'}( \omega(0), d\omega) C(\omega)\leq
3e^{c'\beta' T}\sum_{n\geq |\|\omega_0\|^2}e^{
{_{n}\over^{8R}}} e^{- {_{n}\over^{4R}}}+
C(\omega(0))= C'(\omega(0)).
\label{x22}
\qqq
 Hence, $\lim_{t\to
\infty}P^t(\omega(0),1_E)$ exists, and
 (\ref{con}) with $F(l')
=1$ also follows.

\vs 2mm

Now, consider  (\ref{con}) for a general $F=F(l')$. Write
$F= F-F_0+F_0$, where, by definition, $F_0(l) = F(l (t, {
s}([\frac{t}{2}, t]), 0))$. Then,
\qq
P^t(\omega(0),F)=\int\mu^t_{\omega(0)}(d\Ns)(F-F_0)+
\int\mu^t_{\omega(0)}(d\Ns)F_0.
\label{x150}
\qqq
Let us start with the first term. We write it as
\qq
\int\mu^t_{\omega(0)}(d\Ns)(F-F_0)=
\int\mu^t_{\omega(0)}(d\Ns)(F-F_0)1_\omega+
\int\mu^t_{\omega(0)}(d\Ns)(F-F_0)(1-1_\omega)
\label{x15}
\qqq
where $1_\omega$ is the indicator function of the event
$\|\omega(\frac{t}{2})\|^2> Rt$. By  the probabilistic
estimate (\ref{c5}) the first term may be bounded by
\qq
2\|F\|_\infty P(\|\omega(\frac{t}{2})\|^2> Rt|\omega(0))\leq
C(\omega(0)) \|F\|_\alpha
e^{-ct}
\label{x16}
\qqq
where $\|F\|_\alpha$ is the H\"older norm of
$F$.

For the second term, write it as a sum
\qq
\int\mu^t_{\omega(0)}(d\Ns)(F-F_0)(1-1_\omega)1_D+
\int\mu^t_{\omega(0)}(d\Ns)(F-F_0)(1-1_\omega)(1-1_D)
\label{x20}
\qqq
where $1_D$ is the indicator function of the event
$a\sum_{n=\frac{t}{2}+1}^{t}D_n(\omega)>\frac{\kappa}{2}Rt$.
Using again the probabilistic estimates, we have, by
(\ref{x1}) (with $0$ replaced by $\frac{t}{2}$) and the
constraint $1-1_\omega$, i.e.
$\|\omega(\frac{t}{2})\|^2\leq Rt$, that, for $\kappa$
large:
\qq
|\int\mu^t_{\omega(0)}(d\Ns)(F-F_0)(1-1_\omega)1_D|\leq C
\|F\|_\alpha e^{-ct}.
\label{x17}
\qqq
 For the second term in (\ref{x20}), we
use the fact that $F$ is H\"older continuous:
$$ |F-F_0|\leq \|F\|_\alpha \|l (t, {\bf s}([{_{t}\over^{2}}, t], 0)-
l (t, {\bf s}([0, t],l_0)\|^\alpha,
$$
 and
$$\|l (t, {\bf s}([{_{t}\over^{2}}, t], 0)-
l (t, { \bf s}([0, t],l_0)\|
=\|l (t, { \bf s}([{_{t}\over^{2}}, t], 0)-
l (t, {\bf s}([{_{t}\over^{2}}, t],
l({_{t}\over^{2}}))\|\leq  e^{-cRt},
$$
which follows from (\ref{21}), with $[0, t]$ replaced by
$[{_{t}\over^{2}}, t]$, given that we have here both the
constraint that
$$a\int_{\frac{t}{2}}^t
\|\nabla \omega \|^2\leq
a\sum_{n=\frac{t}{2}+1}^{t}D_n(\omega)\leq\frac{\kappa}{2}Rt,
$$
and that $\| l_1(\frac{t}{2})- l_2(\frac{t}{2})\|^2=
\|l(\frac{t}{2})\|^2\leq
\|\omega(\frac{t}{2})\|^2\leq Rt$. Thus,
\qq
|\int\mu^t_{\omega(0)}(d\Ns)(F-F_0)(1-1_\omega)(1-1_D)|\leq
\|F\|_\alpha e^{-c\alpha Rt}
\label{x18}
\qqq

Altogether, combining (\ref{x15}-\ref{x18}), we get:
\qq
|\int\mu^t_{\omega(0)}(d\Ns)(F-F_0)|\leq C(\omega(0))\|F\|_\alpha
e^{-ct}.
\label{x19}
\qqq
where $c=c(R, \alpha)$.

Returning to (\ref{x150}), we will finish the proof by
bounding
\qq
\int\mu^t_{\omega_0}(d\Ns)F_0-\int\mu^t_{\omega'_0}(d\Ns)F_0.
\non
\qqq
We insert
the expansion (\ref{z1}) in each term and
integrate over ${\bf s}([0, t])$; since
$F_0$ depends only on ${\bf s}([{_{t}\over^{2}}, t])$, we obtain,
in each term of the sum,
a formula like (\ref{PT1}) for the factors
occurring before the first $K_i$ intersecting $[\frac{t}{2}, t]$
(and an expression depending on $F_0$ for the rest).
 Now,
expand the resulting factors $\mu^{n_i}$, after $t_0$, as
above (see
 the arguments leading to (\ref{x13})).
 As before, let $P^t_n(\omega_0, F_0)$
 collect all the terms containing a factor $\mu_0$
(after $t_0$ and before the first $K_i$ intersecting
$[\frac{t}{2}, t]$). Again, $P^t_n({\omega_0},
F_0)=P^t_n({\omega'_0}, F_0)$. Now, for $R^t(\omega_0,
F_0)$, we first bound $F_0$ by its supremum, then bound
each term of the resulting expansion, using (\ref{x25})
for the $\mu$ factors and (\ref{muk}) for the other
factors. The result is
$$
|\int\mu^t_{\omega_0}(d\Ns)F_0-\int\mu^t_{\omega'_0}(d\Ns)F_0|
\leq C(\omega(0))\|F_0\|_\infty
 (1
-\delta')^{t/2T}e^{e^{-cT} {_{t}\over^{T}}},
$$
where the $(1
-\delta')^{t/2T}$ factor comes from the fact
that, in $R$, we have only the factors
$\mu_K$
with $|K|\geq 2T$, $\Delta^m$, $\lambda^p-\mu_0 $, $\Delta\bar
\mu$, $\bar
\mu$, $\lambda^{q}$, with $q<p$,
 appearing during the
time interval $[t_0,t/2]$ and we can therefore use
(\ref{603}), with $t$ replaced by $t/2$ to obtain a lower
bound on the number of $\lambda^p(d\Ns|\Ns')-
\mu_0(d\Ns)$ factors. Combining this with
(\ref{x19}), (\ref{x150}), we obtain (\ref{x130}) with
$1_E$ replaced by $F$. To finish the proof, we can now
use arguments like (\ref{x21}-\ref{x22}) to get
(\ref{con}) in general.\hfill$\Box$
\vs 5mm

The next two Sections will be devoted to the proof of,
respectively,
 Propositions 3 and 4.

\section{Proof of Proposition 3}

Consider the expression
\qq
X(\sigma')\equiv
\int|\Delta^m\mu_{K}(d\sigma|\sigma')|
\label{Xs'}
\qqq
for $K=[\tau_0,\tau]$ a $T$-interval. Let
$\pi=(J_1,\dots,J_{n-m})$ be a partition of $K$ in the
sum (\ref{muKk}) and define also for $i\in [1, m]$
$J_{n-m+i}=[\tau+(i-1)T,\tau+iT]$. Hence the $J_i$,  for
$i\in[1,n]$,  form a partition of the set $\bar K=
[\tau_0,\tau+mT]$. Let $\Nk_i\in \NN^{J_i}$,
$i=1,\dots,n$. Set $\Nk=(\Nk_1,\dots ,\Nk_{n},
)\in\NN^{|K|+mT}$. Finally, let
$\Nj=(j_3,\dots,j_{n-m})$. Then combining the definitions
(\ref{muKk}), (\ref{musigma}), (\ref{muk1}) and
(\ref{Delta}), we can bound
\qq
X(\sigma')\leq
\sum_{\pi\Nk\Nj}
\int
\Big|\prod_{i=1}^{n-m}\mu_{\Nk_i,
j_i}(d\Ns(J_i)|{\bf s}([\tau_{j_i-1},
\tau_{i-1}]))
\prod_{i=n-m+1}^{n}\mu_{\Nk_ii-1}(d\Ns(J_i)|\Ns(J_{i-1}))
\Big|1(\Nk|\Nk')
\label{z2'}
\qqq
where ${\bf s}([\tau_0-T,\tau_0])=\Ns'$ unless $0\in K$, in
which case $\Ns'\equiv \omega(0)$. We also put
\qq
1(\Nk|\Nk')=\prod_{i=0}^{n-m-1}1_{J_iJ{i+1}}(\Nk_i,\Nk_{i+1})
\prod_{i=n-m}^{n-1}|1_{J_iJ{i+1}}(\Nk_i,\Nk_{i+1})-1|
\prod_{i=1}^{n}1_{J_i}(\Nk_i),
\label{1Nk'}
\qqq
with $\Nk_0=\Nk'$ and the sum over $\pi$, $\Nj$,  has the
constraint that the set (\ref{Kalpha}) or (\ref{K'alpha})
is $K$. 
Let $\CI=\{i\;|\; j_i\neq i-1\}\subset
\{3,
\ldots, n-m\}$. For $i\in \CI$, we rewrite $\mu_{\Nk_i,j_i}$ (see
eq. (\ref{mui,j})) as
$$
\mu_{\Nk_i,{j_i}}=
\chi_{\Nk_i{j_i}} g_{i{j_i}}-\chi_{\Nk_i{j_i}+1}
g_{i{j_i}+1}= (\delta_i\chi+\chi_{\Nk_i{j_i}+1}\delta_i
g) g_{i{j_i}}
$$
where
\qq
\delta_i\chi=\chi_{\Nk_i{j_i}}-\chi_{\Nk_i{j_i}+1}
\label{deltachi}
\qqq
and
\qq
\delta_i g=1-{g_{i{j_i}+1}\over g_{i{j_i}}}.
\label{dg}
\qqq
Introducing the  probability measures
\qq
\mu_{\Nj}=
\prod_{i=1}^ng_{ij_i}
\nu_{s(\tau)}^{|K|},
\label{zz}
\qqq
where $j_i\equiv i-1$ if $i>n-m$, we can write (\ref{z2'}) as
\qq
X(\sigma')\leq\sum_{\pi\Nk\Nj}\sum_{A\subset \CI}
\int\prod_{i\in A}|\delta_i\chi|
\prod_{i\in B}\chi_{\Nk_ij_i+1}|\delta_i g|
\prod_{i\in C}\chi_{\Nk_ii-1}\mu_{\Nj}
1(\Nk|\Nk')
\equiv \sum_{\pi\Nk\Nj A}\int \CR_{\Nk\Nj A}\mu_{\Nj}1(\Nk|\Nk')
\label{activity3}
\qqq
where $B=\CI\setminus A$ and $C=\{1,\dots,n\}\setminus
\CI$.

Letting
$$
\delta_if=f_{j_i+1}-
f_{j_i},
$$
(\ref{dg}) can be written as
\qq
\delta_i g=1-e^{\int_{J_i}(\delta_if(t),\gamma^{-1}(ds(t)-
f_{j_i}(t)dt))
-\hf \int_{J_i}(\delta_if(t),\gamma^{-1}\delta_if(t))dt
}\equiv 1-H_i
\label{dg1}
\qqq
and
\qq
\delta_i\chi=
\prod_{\Nn\subset J_i}\phi_{k_n}(D_n(\omega_{j_i}))-
\prod_{\Nn\subset J_i}\phi_{k_n}(D_n(\omega_{j_i+1}))
\label{dg2}
\qqq

We will now undo the Girsanov transformation, i.e. change
variables from $s$ back to $b$. Let $E$ denote the
expectation with respect to the Brownian motion $b$ with
covariance $\gamma$ on the time interval $K$. Then,
\qq
\int\CR_{\Nk\Nj A}\mu_{\Nj}=E\CR_{\Nk\Nj A}
\label{muk2}
\qqq
where $\CR$ is given by the same expression as before,
but the symbols $s$ and $\omega_{j_i}$ have to be
interpreted as follows:
$s$ is the progressively measurable function
of $b$
defined on each interval $J_i$ as the solution of
\qq
ds(t)= f_{j_i}(t)dt+db(t),
\label{fjieq}
\qqq
where $f_j(t)=PF(\omega_{j}(t))$ and $\omega_{j}(t)=
s(t)+l(t,{\bf s}([\tau_{j-1},t]),0)$, with, for $i=1$,
$\Ns([\tau_{j_1-1}, \tau_0])$ replaced by $\Ns' (J_0)$,
which expresses the dependence of (\ref{muk2}) on $\Ns'$;
 $H_i$, defined by (\ref{dg1}), can be written:
\qq
H_i=e^{\int_{J_i}(\delta_if(t),\gamma^{-1}db(t))
-\hf \int_{J_i}(\delta_if(t),\gamma^{-1}\delta_if(t))dt.
}
\label{dg11}
\qqq
We will call the $\omega_{j_i}(t)$ collectively by
\qq
\omega_{\Nj}(t)=\omega_{j_i}(t)\;\;{\rm for}\;\;t\in J_i,
\label{41}
\qqq
and reserve the notation $\omega(t)$ for the solution of
the Navier Stokes equation (\ref{ee}) with given $b(\bar
K)$ and with initial condition
$\omega(\tau_0)=(s(\tau_0), l(\tau_0, {\bf s}'(J_0),0))$
determined by the $\Ns'$ in (\ref{z2'}).

\vs 2mm

\no {\bf Remark}. $\omega_{\Nj}(t)$ is {\it not} a
solution of (\ref{ee}) on the interval $\bar K$
with initial condition given at time $\tau$. On
each interval $J_i$ it solves (\ref{ee}) but
when moving to the next interval the $l$-part
is possibly set equal to zero, depending on $\Nj$.

\vs 2mm

The following Proposition contains the key bounds needed
to estimate (\ref{muk2}).

\vspace*{5mm}

\no{\bf Proposition 5.} {\it Let $b$ belong to the support
of $\CR_{\Nk\Nj A}$ in (\ref{muk2}). Then,
there exists a constant $c$ such that
\qq
||\delta_if(t)||\leq
e^{-c\kappa R {\rm dist}(J_i,J_{j_i})}\equiv \epsilon_i
\label{lemma12}
\qqq
and
\qq
\prod_{i\in A}|\delta_i\chi|
\prod_{i\in B}\chi_{\Nk_ij_i+1}
\prod_{i\in C}\chi_{\Nk_ii-1}\leq \prod_{i\in A}|J_i|\epsilon_i
1_{{\Nk}}(\omega)
\label{lemma13}
\qqq
where $1_{{\Nk}}(\omega)$ is the indicator function of
the set of $b\in C(\bar K,H_s)$ such that, for all $\Nn\subset \bar K$,
we have
\qq
D_n(\omega)\in [2^{k_n-1}R, 2^{k_n+3}R],\, for\,k_n\neq 0;\,
D_n(\omega)\in [0, 5R],\, for\, k_n=0
\label{42}
\qqq
and $\omega$ is the solution of the Navier-Stokes equation explained above.
}

\vspace*{5mm}

Let $\eta_i(t)$ be the indicator function of the event
that $\delta_if(t)$ satisfies the bound (\ref{lemma12}).
$\eta_i(t)$ is progressively measurable. Since $\eta_i=1$
on the support of the summand in (\ref{activity3}),  we
may replace $\delta_if(t)$ there by
$\eta_i(t)\delta_if(t)$. Denote $H_i$, defined in
(\ref{dg1}), after this replacement, by $\bar H_i$. We
have, using (\ref{lemma13}),
\qq
E\CR_{\Nk\Nj A}\leq
\prod_{i\in A}|J_i| \epsilon_i
E\left(\prod_{i\in B}(1-\bar H_i)
1_{\Nk}\right)
\non
\qqq
and inserting this to (\ref{activity3})
\qq
X(\sigma')\leq\sum_{\pi\Nj A}
\prod_{i\in A} |J_i|\epsilon_i
E(\prod_{i\in B}(1-\bar H_i)1_\pi(\omega)).
\label{61}
\qqq
where
\qq
1_\pi(\omega)=
\sum_{\Nk}
1_{\Nk}(\omega)1(\Nk|\Nk').
\label{1pi}
\qqq
The expectation in (\ref{61}) is bounded using Schwarz'
inequality by
\qq
(E\prod_{i\in B}(1-\bar H_i)^2)^\hf
(E 1_{\pi}^2)^\hf
\label{xx}
\qqq
To estimate the first square root
renumber the intervals $J_i$ for $i\in B$
as $J_1,\dots,J_b$, $J_i=[\sigma'_i,\sigma_i]$
with $\sigma_1>\sigma'_1 \geq \sigma_2\dots$. Denote expectations in
the Brownian filtration $\CF_\tau$ by $E_{\tau}$.
Then
\qq
E\prod_{i\in B}(1-{\bar H}_i)^2=
E_{\sigma'_1}\Bigl(E_{\sigma_1}((1-{\bar
H}_{1})^2\;|\;\CF_{\sigma'_1})\prod_{i>1}
(1-{\bar H}_i)^2\Bigr).
\label{bbb}
\qqq
Expanding $(1-{\bar H}_i)^2=1-2{\bar H}_i +{\bar H}_i^2$, we first
bound from below, using (\ref{lemma12}) and Jensen's inequality,
\qq
E_{\sigma_1}({\bar H}_{1}
|\CF_{\sigma'_1})\geq\exp(-\hf |J_1|\epsilon_1^2\rho^{-1}).
\label{s20}
\qqq
For an upper bound for the expectation of ${\bar H}_1^2$,
we use

\vspace*{2mm}

\no{\bf Lemma 7.1.} {\it Let $\zeta(t)\in C([0,t], H_s)$ be
progressively
measurable. Then
\qq
Ee^{\int_0^t (\zeta,\gamma^{-1}db)
+\lambda\int_0^t(\zeta,\gamma^{-1}\zeta) dt}\leq e^{2(1+\lambda)t
||\zeta||^2\rho^{-1}}
\label{novikov}
\qqq
where $||\zeta||=\sup_\tau||\zeta(\tau)||_2$.}

\vs 2mm

\no{\bf Proof}. This is just a Novikov bound: we bound the LHS,
using Schwarz' inequality, by
$$
(Ee^{\int_0^t (2\zeta,\gamma^{-1}db)
-2\int_0^t(\zeta,\gamma^{-1}\zeta dt)})^\hf
(Ee^{2(1+\lambda)\int_0^t(\zeta,\gamma^{-1}\zeta) dt})^\hf
$$
and note that the expression inside the first square root
is the expectation of a martingale and equals one.\hfill$\Box$

\vs 2mm

Applying Lemma 7.1 to $\zeta=2\eta_i\delta_if$ and
$\lambda=-{_1\over^4}$
we obtain
\qq
E_{\sigma_1}({\bar H}_{1}^2|\CF_{\sigma'_1})
\leq \exp(3 |J_1|\epsilon_1^2\rho^{-1}).
\label{no}
\qqq
This and (\ref{s20}) imply
$$
E_{\sigma_1}((1-{\bar H}_{1})^2\;|\;\CF_{\sigma'_1})\leq
C|J_1|\epsilon_1^2\rho^{-1}\exp(C|J_1|\epsilon_1^2\rho^{-1}).
$$
Iterating the argument, we arrive at
\qq
E\prod_{i\in B}(1-{\bar H}_i)^2\leq
\prod_{i\in B}C|J_i|\epsilon_i^2\rho^{-1}
\exp(C|J_i|\epsilon_i^2\rho^{-1}).
\label{barg}
\qqq
Since  ${\rm dist}(J_i,J_{j_i})\geq T$, by choosing
$T>T(\rho)$ we may bound the $i$:th factor in
(\ref{barg}) by $\epsilon_i$  if $J_i$ is small (so that
$|J_i|=T$) and, by $\epsilon_i e^{\delta|J_i|}$ if $J_i$
is large,
 where $\delta$ can be made arbitarily
small by increasing $T$.
Thus, we may combine
(\ref{61}), (\ref{xx}) and (\ref{barg}), to get
\qq
X(\sigma')\leq\sum_{\pi\Nj A}
\prod_{i\in A\cup B}e^{-c\kappa R{\rm dist}(J_i,J_{j_i})}
\prod_{|J_i|>T}e^{\delta|J_i|}
(E1_{\pi}^2)^\hf.
\non
\qqq
Writing $c=2c_1$, the sums over $\Nj$ and $A$ are
controlled by
$$
\sum_{\Nj}\prod_{i\in A\cup B}e^{-c_1\kappa R{\rm dist}(J_i,J_{j_i})}\leq
e^{-c_2\kappa RT|A\cup B|}
$$
(since  ${\rm dist}(J_i,J_{j_i})\geq T$) and
$$
\sum_{ A\subset K}e^{-c''\kappa RT|A|}<2^{|K|}
$$
and the last expectation by

\vspace*{3mm}

\no{\bf Lemma 7.2.}
{\it Under the assumptions of Proposition 3,
\qq
E1_{\pi}^2\leq C_K(\omega(0))C^{|K|+mT}e^{-c\beta' mT}
\prod_{J_i\; large}e^{-c\beta'|J_i|}
\non
\qqq
with  $C_K(\omega(0))$ has the same form
as in Proposition 3 (with another $c'$).}

\vspace*{3mm}

We are thus left with the bound
\qq
X(\sigma')\leq C_K(\omega(0)) C^{|K|+mT}e^{-c \beta' mT}
\sum_{\pi}
\prod_{J_i\;{\rm large}}e^{-c\beta'|J_i|}
\sup_{\Nj}\prod_{i\in \CI}e^{-c_1\kappa R{\rm dist}(J_i,J_{j_i})}.
\label{xxxx}
\qqq
where we recall that $\CI=A\cup B=  \{i\;|\; j_i\neq
i-1\}$.

Let first $0\notin K$. Then $K$ is the union of the sets
on the LHS of (\ref{K'alpha}). Each small $J$ is either a
subset of $[\tau_{j_i-1},\tau_i]$ or a $J_{i+1}$ for
$J_i$ large. Thus
 the summand in (\ref{xxxx}) is smaller than $e^{-c\beta'|K|}$,
 for $\kappa R\geq \beta'$.
For $0\in K$, we have a similar bound, except that
$J_1,J_2$ may be small and not in any
$[\tau_{j_i-1},\tau_i]$ so that $|K|$ is replaced by
$|K|-2T$; but the $2T$ may be absorbed to the $c'\beta' T$
in $C_K(\omega_0)$ (see (\ref{CK})). The sum over $\pi$
is a sum over partitions of $K$ into T-intervals (with
labels for the  intervals of length $T$), and thus is bounded by
$C^{|K|}$. Thus the claim (\ref{muk}) follows for
$\beta'$ large enough.\hfill$\Box$

\vs 5mm

\no {\bf Proof of Proposition 5}. Let us start
with the proof of (\ref{lemma13}). For that, we need to
have  a bound on the difference
$|D_n(\omega_{j_i})-D_n(\omega_{_{j_{i+1}}})|$, which is
the difference between the arguments of the two $\chi$
functions in (\ref{deltachi}) (see (\ref{chik1})). For
that, we need some lemmas. Remember the definition
$\omega_j(t)=\omega(t,\Ns([\tau_{j-1},t]),0)$. We have

\vs 2mm

\no{\bf Lemma 7.3.} {\it Let $n>m\geq\tau_i$, $i>j$. Then
\qq
|D_n(\omega_i)-D_n(\omega_{j})| \leq
e^{- \kappa R(n-m-1)
+ a\sum_{p=m+1}^n D_p(\omega_i)}(\| \delta l(m) \|
+\| \delta l(m) \|^2)
\label{400}
\qqq
where} $\delta l=l_i-l_{j}$.

\vs 2mm

\no{\bf Proof}. By definition,
\qq
|D_n(\omega_i)-D_n(\omega_{j})| \leq \hf |\sup_t ||l_i(t)||^2-
\sup_t||l_{j}(t)||^2|+|\int_\Nn\| \Nna l_i(t)||^2dt-\int_\Nn\|
\Nna l_{j}(t)||^2dt|
\label{402}
\qqq
The second term is bounded by
\qq
\int_\Nn\| \Nna\delta l(t) \|(2\| \Nna
\omega_i(t) \|+\| \Nna \delta l(t) \|)dt,
\label{s40}
\qqq
Remembering the calculation in Proposition 1,
(\ref{a9}), (\ref{a10}), we have:
\qq
\int_\Nn\| \Nna \delta l (t)\|^2dt\leq  \| \delta l(n-1) \|^2
+a\int_\Nn\| \delta l (t)\|^2\| \Nna
\omega_i (t)\|^2 dt.
\label{s30}
\qqq
Using (\ref{21}), the second term is bounded by
\qq
a\int_\Nn \| \Nna \omega_i(t) \|^2e^{2a\int_{n-1}^t\| \Nna
\omega_i(\tau) \|^2}
\| \delta l(n-1) \|^2\leq (e^{2aD_n(\omega_i)}-1)\| \delta l(n-1) \|^2
\label{s6}
\qqq
which, together with (\ref{s30}), yields
\qq
\int_\Nn\| \Nna \delta l (t)\|^2dt\leq
e^{2aD_n(\omega_i)}\| \delta l(n-1) \|^2 .
\non
\qqq
Now, using this for the second term on the RHS of (\ref{s40}), and
Schwarz' inequality to bound the first one, we get
\qq
(\ref{s40}) \leq
2D_n(\omega_i)^\hf e^{ a D_n(\omega_i)}\| \delta l(n-1) \|+e^{2aD_n(\omega_i)}\|
\delta l(n-1)
\|^2,
\label{4020}
\qqq
since $\int_\Nn\| \Nna
\omega_i(t) \|^2 dt \leq D_n(\omega_i)$.

For the first term of (\ref{402}), use $||l_{j}(t)||^2=||l_i(t)||^2+
2(\delta l(t),l_i(t))+||\delta l(t)||^2$ to bound it by
$2\sup_{t \in \Nn}|(\delta l(t),l_i(t))| + \sup_{t \in \Nn }||\delta l(t)||^2$,
which, by Schwarz' inequality, $\sup_{t \in \Nn}\|l_i(t))\|\leq D_n(\omega_i)$,
 and (\ref{21}),
leads again to the bound (\ref{4020}). This yields our claim if we use (\ref{21})
to bound $\| \delta l(n-1) \|$.\hfill$\Box$

\vs 2mm

 To be able to apply this lemma, we need
to bound $D_p(\omega_i)$ in the exponent of (\ref{400});
Note that the functions $\chi$ in (\ref{activity3}) put
constraints (to be in the interval
$[2^{k_p}R,2^{k_p+2}R]$, for $p\in J_i$), but the latter
apply to $D_p(\omega_{j_i})$ or $D_p(\omega_{j_i+1})$,
 not directly to $D_p(\omega_i)$.
So, we need to compare those different $D_p$'s. This will
be done in Lemma 7.5 below, whose proof will use

\vs 2mm

\no{\bf Lemma 7.4}
{\it Suppose  that $\Nu\subset J_q$ and $p\leq q-1$. Then
\qq
\sum_{\tau_{p-1}< l\leq u}2^{k_l}\leq 2\beta(u-\tau_{p-1}) .
\label{large11}
\qqq
}

\vs 2mm

\no{\bf Proof.}  Let $q\leq n-m$. Then $L=[\tau_{p-1},
u]$ cannot satisfy (\ref{large}) (otherwise $[\tau_{p-1}, u]$ would
be inside the same large interval) and so,
\qq
\gamma_L\leq \beta(u
-\tau_{p-1})
\label{2ss1}
\qqq
and  the claim is true. So, suppose that $n-m<q$.
The interval  $L=[\tau_{p-1}, \tau_{n-m-1}+1]$ (which is empty if $n-m<p$)
cannot satisfy (\ref{large}) either and so,
\qq
\gamma_L\leq\beta(|L|+1).
\label{2ss2}
\qqq
The intervals $J_i$ are small
if $i\geq n-m$ ($J_{n-m}$ is small since the last interval in $K$
is small) and
thus $\gamma_{J_i}\leq
\beta|J_i|=\beta (\tau_i-\tau_{i-1})$. Hence, (\ref{2ss2}) holds
for $L=[\tau_{p-1},\tau_{q-1}]$.
 Altogether, we get
\qq
\sum_{\tau_{p-1}< l\leq u}2^{k_l}\leq \beta(\tau_{q-1}
-\tau_{p-1}+1)+\sum_{\tau_{q-1}< l\leq u}2^{k_l}.
\label{2ss3}
\qqq
The last term in
(\ref{2ss3}) is bounded by $\beta\max\{\hf
T,u-\tau_{q-1}\}\leq \beta (u-\tau_{q-1} +\hf T)$,
 since the small $J_q$ cannot contain
an $L$ satisfying (\ref{large}). Hence, (\ref{2ss2}) is
bounded by $\beta(u-\tau_{p-1}+1+\hf T)$ which in turn is
bounded by (\ref{large11}) since $u-\tau_{p-1}\geq T$.
\hfill$\Box$

\vs 2mm

\no{\bf Lemma 7.5.} {\it Let $s$ be in the support
of the measure in the summand of (\ref{activity3}).
Let $q\geq 2$ if $0\notin K$, $q\geq 3$ if $0\in K$.
Let $\Nn\subset J_q$ and $i,j\leq q-1$. Then,
\qq
|D_n(\omega_i)-D_n(\omega_{j})| \leq
e^{-{_\kappa\over^2}TR}
\label{404}
\qqq
}

\vs 2mm

\no{\bf Proof}. We perform an induction in $q$.
Suppose that the claim holds up to $q-1$. Let $\Nn
\subset J_{q}$. In eq. (\ref{activity3}), because of the
functions $\chi$, the measure is supported
 on configurations where, for each $p$,
either, $\forall \Nm \subset J_p$, $D_m(\omega_{j_p})$ is
constrained to be in the interval
$[2^{k_m}R,2^{k_m+2}R]$, or, $\forall \Nm \subset J_p$,
$D_m(\omega_{j_p+1})$ is constrained to be in that
interval; remember that both ${j_p}$ and ${j_p+1}$ are
less than or equal to  $p-1$. Let us consider, for each
$p$, an arbitrary choice between ${j_p}$ and ${j_p+1}$
and call it $j^p$. Thus, to repeat,
\qq
D_m(\omega_{j^p})\in [2^{k_m}R,2^{k_m+2}R]\;{\rm for}\;\Nm
\subset J_p\;
 ,\; {j^p}\leq p-1.
\label{j^p}
\qqq
Since the support of the measure in  (\ref{activity3})
contains only configurations such that (\ref{j^p}) holds
for some choice of the function $j^p$, it is enough to
bound $|D_n(\omega_i)-D_n(\omega_{j^q})|$ and
$|D_n(\omega_j)-D_n(\omega_{j^q})|$ for an arbitrary
function $j^p$, assuming that (\ref{j^p}) holds.

>From Lemma 7.3, we get, for $\Nn\subset J_q$,
\qq
|D_n(\omega_i)-D_n(\omega_{j^q})| \leq
e^{- \kappa R(n-\tau_{q-1}-1)
+ a\sum_{m=\tau_{q-1}+1}^n D_m(\omega_{j^q})}(\| \delta l(\tau_{q-1}) \|
+\| \delta l(\tau_{q-1}) \|^2)
\label{405}
\qqq
where $\delta l=l_i-l_{j^q}$. We need to estimate $\|
\delta l(\tau_{q-1}) \|$. For this to be nonzero, $i$ and
$j^q$ cannot be equal, and they are both less than or
equal to $q-1$. So, let us say that $i<q-1$. Then, by
Proposition 1,
\qq
\| \delta l(\tau_{q-1}) \|\leq e^{-\kappa R(\tau_{q-1}-
\tau_{q-2})
+ a\sum_{m=\tau_{q-2}+1}^{\tau_{q-1}} D_m(\omega_{i})}\| \delta
l(\tau_{q-2}) \| .
\label{406}
\qqq

Now, use (\ref{j^p}) for $p=q$, to get that, in
(\ref{405}), $D_m(\omega_{j^{q}})\leq 2^{k_m+2}R$;   in
(\ref{406}), we note that, since $\Nm \subset J_{q-1}$,
and since both $i$ and $j^{q-1}$ are less than or equal
to $q-2$, we have, by the induction hyphothesis,
$$
|D_m(\omega_i)-D_m(\omega_{j^{q-1}})| \leq
e^{-{_\kappa\over^2}TR}
$$
 and so, $D_m(\omega_{i})\leq 2^{k_m+2}R+
e^{-{_\kappa\over^2}TR}<2^{k_m+3}R$.
Since we shall show below that $\| \delta l(\tau_{q-1}) \| \leq 1$ which
implies
$\| \delta l(\tau_{q-1}) \|
+\| \delta l(\tau_{q-1}) \|^2\leq 2 \| \delta l(\tau_{q-1}) \|$  we obtain,
by combining (\ref{405}) and (\ref{406}),
\qq
(\ref{405})\leq 2e^{- \kappa R(n-\tau_{q-2}-1)
+ aR\sum_{m=\tau_{q-2}+1}^n 2^{k_m+3}}\|
\delta l(\tau_{q-2}) \| .
\label{4061}
\qqq
Now, remember that $i\leq q-2$. If $j^q=q-1$, then $l_{j^q}(\tau_{q-2})=0$ and
$$
\|\delta l(\tau_{q-2}) \|=\|l_i(\tau_{q-2}) \|\leq
\|l_i(\tau_{q-2})- l_{j^{q-1}}(\tau_{q-2})\|+
\|l_{j^{q-1}}(\tau_{q-2})\|.
$$
Since both $i$ and $j^{q-1}$ are less than or equal to
$q-2$, the first term is bounded by
$e^{-{_\kappa\over^2}TR}$, using the inductive
hyphothesis. For the second term, we use
$\|l_{j^{q-1}}(\tau_{q-2})\|^2\leq
\|\omega_{j^{q-1}}(\tau_{q-2})\|^2\leq 8\beta RT$, see
(\ref{s60}).
Thus, altogether,
$$
\|
\delta l(\tau_{q-2})\| \leq (C\beta RT)^\hf.
$$
If $j^q<q-1$ then, by induction, $\|
\delta l(\tau_{q-2})\| \leq e^{-{_\kappa\over^2}TR}$.
By Lemma 7.4 , eq. (\ref{large11})
$$
\sum_{m=\tau_{q-2}+1}^n 2^{k_m}\leq 2\beta (n- \tau_{q-2}).
$$
Combining these observations,
$$
(\ref{405})\leq e^{-(\kappa-C\beta)TR} (C\beta RT)^\hf
$$
since $n-\tau_{q-2}\geq
T$.
The same result holds for $i$ replaced by $j$
and hence also for $D_n(\omega_i)-D_n(\omega_j)$. Thus,
the inductive claim (\ref{404}) follows  provided $\kappa>\kappa(\beta)$.

To start the induction,
we need to distinguish the cases $0\in K$ and $0\notin K$.
Start with the latter. Let $J_0=[\tau_0-T,\tau_0]$.
We have $q=2$, $\{i,j\}=\{0,1\}$ (since the case $i=j$ is trivial).
Now $j^{1}$ must be $0$ or $1$, i.e. we may assume $\omega_i-\omega_j
= \omega_i-\omega_{j^{1}}$. Since
for $\Nm \subset J_1$, $D_m(\omega_{j^{1}})\leq
2^{k_m+2}R$, Proposition 1 implies, with  $\delta l=l_i-l_j$,
\qq
\| \delta l(\tau_{1}) \|\leq
e^{- \kappa R(\tau_1-\tau_{0}-1)
+ aR\sum_{m=\tau_{0}+1}^{\tau_1} 2^{k_m+2}}
\| \delta
l(\tau_{0}) \|.
\label{4062}
\qqq

Thus we need to estimate $\|\delta
l(\tau_{0}) \|$. Since $l_1(\tau_0)=0$, this equals
$\|l_0(\tau_0)\|$. But this in turn is, by
(\ref{s60}), bounded by $\|\omega_0(\tau_0)\| \leq (8\beta RT)^{\hf}$. We may now
proceed as above, using (\ref{405}) to obtain
 the claim for $q=2$.

Let finally $0\in K$. Now we start the induction from $q=3$,
and may assume $i=1$ and $j=2$.
 We should also remember that
now, $\omega_1(t)=(s(t), l(t,\Ns([0,t]),l(0))$ for $t\in
[0,\tau_2]$ where $l(0)=(1-P)\omega(0)$. By contrast,
$\omega_2(t)=(s(t), l(t,\Ns([\tau_1,t]),0)$ for $t\in
[\tau_1,\tau_2]$. For $\Nm \subset J_1\cup J_2$ we have
$D_m(\omega_1)\leq 2^{k_m+2}R$ since no decoupling was
done on those intervals. Proceding as in the previous
case we obtain
\qq
\| \delta l(\tau_{2}) \|\leq
e^{- \kappa R(\tau_2-\tau_{1}-1) +
aR\sum_{m=\tau_{1}+1}^{\tau_2} 2^{k_m+2}}
\|l_1(\tau_{1}) \|.
\label{4063}
\qqq
Now, again by (\ref{s60}), $\| l_1(\tau_{1}) \|^2\leq \|
\omega_1(\tau_{1}) \|^2 \leq 8\beta RT$.
We complete the proof for $q=3$ again
by using  (\ref{405}).\hfill $\Box$

\vs 3mm

Returning to the proof of Proposition 5 and, combining the
Lemmas 7.3 and 7.5, we deduce that, for $n\in J_i$,
\qq
|D_n(\omega_{j_i})-D_n(\omega_{j_i+1})| \leq e^{- \kappa
R(n-\tau_{j_i}-1) + aR\sum_{m=\tau_{j_i}+1}^n
2^{k_p+3}}(\| l_{j_i}(\tau_{j_i}) \| +\|
l_{j_i}(\tau_{j_i}) \|^2).
\label{4000}
\qqq
By (\ref{large11}) the exponent is bounded from above by
$-c\kappa R{\rm dist} (J_i,J_{j_i})$, for $\kappa \geq
\kappa (\beta)$. Since ${\rm dist} (J_i,J_{j_i})\geq T$ and $\|
l_{j_i}(\tau_{j_i})
\|^2\leq CR\beta T$ (using
(\ref{s60}) and (\ref{404})), (\ref{4000}) is bounded by
$\epsilon_i$. Then,
$$
|\delta_i\chi|=|\prod_{\Nn\subset J_i}\phi_{k}(D_n(\omega_{j_i}))-
\prod_{\Nn\subset J_i}\phi_{k}(D_n(\omega_{j_i+1}))|\leq |J_i|\epsilon_i
1_\Nk|_{J_i}
$$
since we may choose $\phi_k$ such that its derivative is
uniformly bounded in $k$. We also used the fact that
$\phi_k$ is supported on $[2^k,2^{k+2}]$ combined with
Lemma 7.5 to bound by $1_\Nk$ that has a larger support
(the latter is much larger than what is needed, but our
choice is notationally convenient). Similarily, Lemma 7.5
allows us to bound $\chi_{\Nk_i i-1}$ and $\chi_{\Nk_i
j_i+1}$ by $1_\Nk$. These observations lead to
(\ref{lemma13}).

The bound (\ref{lemma12}) follows
 from (\ref{a13}) below, using (\ref{21}) to
bound $\| \delta l \|$,   $\sup _{t\in \Nn}\| \omega(t)
\|
\leq (2D_n)^{\hf}$, and Lemmas 7.4, 7.5 to bound the exponent
in (\ref{21}).
\vs 3mm

\noindent {\bf Lemma 7.6}.
{\it Let $f(\omega)=PF(\omega)$ and $\omega=s+l$,
$\omega'=s+l'$. Then,
\qq
\|f (\omega) - f (\omega')\| \leq
C(R)(2\| \omega \| \| \delta l \| + \|
\delta l \|^2)
\label{a13}
\qqq
with $\delta l = l - l'$. Moreover}
\qq
\| f (\omega) \| \leq C(R) ( \| \omega \|
+ \| \omega \|^2)
\label{a14}
\qqq

\noindent
{\bf Proof}. We have
\qq
|f_k (\omega) - f_k (\omega')|
\leq \sum_p |\omega_{k-p} \omega_{p} -
\omega'_{\kappa-p} \omega'_{p} | {|k|\over |p|}
\non
\qqq
which, since $|k| \leq \sqrt{\kappa R}$ is bounded by
\qq
\sqrt{\kappa R} \sum_p | s_{k-p}
\delta l_p + s_p \delta l_{k-p} + l_p
l_{k-p} - l'_p l'_{k-p}|.
\label{a15}
\qqq
Writing $l_p l_{k-p} - l'_p l'_{k-p} =
l_p \delta l_{k-p} + l_{k-p} \delta
l_p - \delta l_p \delta l_{k-p}$ and using Schwarz'
inequality,  we get
\qq
(\ref{a15}) \leq \sqrt{\kappa R} (2 \| \omega \| \| \delta l
\| \| + \|\delta l \|^2)
\non
\qqq
which proves (\ref{a13}), since $f_k \neq 0$ only for
$k\leq \kappa R$.  The proof of
(\ref{a14}) is similar.
\hfill$\Box$

To finish this Section, we have only to give the
\vs 3mm

\no{\bf Proof of Lemma 7.2}. From (\ref{42}) we see that
a given $\omega$ can
belong to the support of at most $5^{|\bar K|}$ different $1_\Nk$.
Furthermore, if $1_\pi(\omega)\neq
0$ then $\omega$ must satisfy the following conditions (remember that
$\beta\geq C \beta'$):

\no a. For each $J_i$, $i\leq n-m$ such that $J_i$ is large
 and $0\notin J_i$, using (\ref{small}) in
Lemma 4.1. and (\ref{42}), we get
\qq
\|\omega(\tau_{i-1})\|^2\leq 16\beta'TR,
\label{l71}
\qqq
 (since $J_{i-1}$ is
small), and, writing $J_i=[\tau_{i-1},\tau_{i-1}+T]
\cup \bar J_i$, either   $[\tau_{i-1},\tau_{i-1}+T]\subset
J_i''$ (see Lemma 4.1, b,
for the definition of $J''$) and
so
\qq
 D_{\tau_{i-1}+T}\geq \hf\beta'TR \;\;{\rm and}
\;\;\sum_{\Nn\subset \bar J_{i}}D_n(\omega) > \hf{\beta'}R |\bar J_{i}|,
\label{l72}
\qqq
(where the second bound always hold for large intervals
since we have $\beta'$ on the RHS; note, however, that
$\bar J_{i}$ could be empty) or
$J_i'\cap[\tau_{i-1},\tau_{i-1}+T]\neq \emptyset$ and so,
in particular, $J_i'\neq \emptyset$ and by (\ref{large2})
and (\ref{42}),
\qq
\sum_{\Nn\subset J_{i}}D_n(\omega) > {_{1}\over^{8}}\beta RT+
\hf{\beta'}R (|J_{i}|-T).
\label{l73}
\qqq
Let $l_i$ be the event (\ref{l72})  and $L_i$ the event (\ref{l73}).

\vs 2mm

\no b. For $0\in J_1$, if $J_1$ is large, then
\qq
\sum_{\Nn\subset J_1}D_n> \hf\beta'R|J_1|.
\label{l80}
\qqq
Let $l_1$ be this event.

\vs 2mm

\no c.
For the set $K'=\bar K\setminus K=[\tau_{n-m},\tau_n]$,
(\ref{l71}) holds for $i-1=n-m$, because, by
construction, the last interval in $K$ is small and
\qq
\sum_{\Nn\subset K'}D_n(\omega) > {_{1}\over^{8}}{\beta}R(m-1)T
\label{l74}
\qqq
(since nearest neighbour $J_i$'s have to be intersected by $L$ with
$\gamma_L>\hf T$). Let $L'$ be the event (\ref{l74}).

\vs 2mm

Let $B$ be the ball in $H$ of radius $16\beta'RT$
 and define
\qq
&&\eta_i=\sup_{\omega(\tau_{i-1})\in B}
P(l_i|\;\omega(\tau_{i-1}))
\non\\
&&\varepsilon_i=\sup_{\omega(\tau_{i-1})\in B}
P(L_i|\;\omega(\tau_{i-1}))
\non\\
&&\epsilon'=\sup_{\omega(\tau_{n-m})\in B}
P(L'|\;\omega(\tau_{n-m}))
\non
\qqq
and in the case of $0\in K$,
$$
\eta_1=P(l_1|\omega(0))
$$
Then we have, for $0\notin K$,
\qq
E1_\pi^2\leq 5^{2|\bar K|}\epsilon'\prod_{J_i\;{\rm
large}} (\eta_i+
\varepsilon_i)
\non
\qqq
and, if $0\in K$, we have $\eta_1$ for $i=1$ replacing
$\eta_1+
\varepsilon_1$. We estimate the $\epsilon$'s and $\eta$ using
Proposition 2.

For $\eta_i$, $0\notin J_i$, apply Proposition 2 with $0$
replaced by $\tau_{i-1}$ (where we use (\ref{l71})) and
$t$ by $\tau_{i-1}+T-1$, and (\ref{l72}):
\qq
\eta_i\leq e^{c_1e^{-cT}\beta'T-c_2\beta'|J_i|}\leq e^{-c\beta'|J_i|},
\non
\qqq
for $T$ large, using also $|J_i|= |\bar J_i| +T$;

For $\varepsilon_i$, Proposition 2, with $0$ replaced by
$\tau_{i-1}$ and (\ref{l73}) give:
\qq
\varepsilon_i\leq e^{c'\beta'T-c_2\beta T-c_3\beta'| J_i|}\leq
e^{-c\beta'|J_i|}
\non
\qqq
which holds for $\beta>C\beta'$; for $\epsilon'$,
Proposition 2, with $0$  replaced by $\tau_{n-m}$, and
(\ref{l74}) give
\qq
\epsilon'\leq e^{c\beta'T-c'\beta (m-1)T}\leq e^{-c\beta (m-1)T}
\label{l75}
\qqq
using $\beta>C\beta'$, and provided $m>1$. For $0\in
J_1$, $J_1$ large, Proposition 2 and (\ref{l80}) give
$$
\eta_1\leq \min\{e^{{_{c}\over^{R}}||\omega(0)||^2-c'\beta'|J_1|},1\}
\leq e^{{_{\delta}\over^{R}}||\omega(0)||^2-c(\delta)\beta'|J_1|}
$$
for any $c>\delta>0$ with $c(\delta)=\delta
{_{c'}\over^{c}}$ (write $c=c-\delta +\delta$, and use the fact that
$\frac{c-\delta}{R}||\omega(0)||^2 \leq (c'-c(\delta))\beta'|J_1|$, whenever
${_{c}\over^{R}}||\omega(0)||^2-c'\beta'|J_1|\leq 0$). We take
$\delta={_{1}\over^{8}}$ (we can always assume that $c$ is larger than that).
Hence, altogether, if $0\in K$,
\qq
E1_\pi^2\leq e^{{_{1}\over^{8R}}||\omega(0)||^2}e^{c\beta'
T} e^{c|
\bar K|}e^{-c\beta'(\sum_{J_i\;{\rm large}} |J_i|+m T)}
\non
\qqq
where $e^{c\beta' T}$ allows to replace $m-1$ by $m$.
Finally, if $0\notin K$,
\qq
E1_\pi^2\leq e^{c| \bar K|} e^{-c\beta' (\sum_{J_i\;{\rm
large}} |J_i|+(m-1) T)}
\non
\qqq

These inequalities give the
claim (since $|K'|= |K| +mT$) except in one case:
 no large $J_i$, $0\notin K$ and $m=1$. In that case,
$J_{n-m+1}$ and $J_{n-m}$ and $J_{n-m-1}$ are all small
($J_{n-m-1}$ is included in $K$, unless $|K|=T$, in which
case the supremum in (\ref{muk}) is taken over $\Ns' \in
C_s$, with $J_0$ in (\ref{Cs}) equal to $J_{n-m-1}$).
Hence, (\ref{muk}) holds for $i-1=n-m-1$.  We may then
apply Proposition 2 with $0$  replaced by $\tau_{n-m-1}$,
use the fact that $m=1$ means that there is an interval
$L$, where (\ref{con1}) is violated, intersecting both
$J_{n-m+1}$ and $J_{n-m}$,
 and get (\ref{l75}) with $m-1$ replaced by $1(=m)$.
\hfill$\Box$

\section{Markov chain estimates.}

 The goal of this Section is to prove Proposition 4.
Although $\lambda (d\Ns|{\bf s'})$ defined in
(\ref{muss'}) does not define a Markov chain, because of
the indicator function $\sum_\Nk\chi_\Nk(\Ns, \Ns')$, it is
close to one, at least up to the time $p$ in which we are
interested, and the proof will be based essentially on
Markov chain ideas. To see how close $\lambda$ is to a
Markov chain, compare it with
$P(d{\bf s}|{\bf s'})=g_J(\omega)\nu^T_{s(\tau)}(d\Ns)$ (see (\ref{Pss'})),
which is thus like $\lambda$, but without the $\sum_\Nk\chi_\Nk(\Ns, \Ns')$;
the function $1-\sum_\Nk\chi_\Nk(\Ns, \Ns')$ is supported on $\Nk$'s such that
$J$ is  a large interval. For ${\bf s'} \in C_s$, we have
$\|\omega (\tau)\|^2\leq 2\beta'TR$ and we can use Proposition 2
to show
that there exists a $c>0$, such
that, $\forall {\bf s'} \in C_s$, $\forall B \subset
C_s$ (note that the support of $\lambda$ is included in $C_s$),
\qq
|\lambda (B|{\bf s'}) -P(B|{\bf s'})| \leq e^{-cT},
\label{p2}
\qqq
and
\qq
P(C_s|{\bf s'}) \geq 1-e^{-cT};
\label{p3}
\qqq
Indeed, if $J$ is large, either there is an interval $L\subset J$
where (\ref{large}) holds, and we use (\ref{x1}) for that
interval, with $0$ replaced by $\tau$, $\|\omega (\tau)\|^2\leq 2\beta'TR$
and $\beta\geq C \beta'$. Or  (\ref{large'}) holds, i.e.
$D_{\tau+T}\geq \beta'TR$, and we can use (\ref{x1}) with
$0$ replaced by $\tau$ and $t=t'-1$ replaced by $\tau+T$.
Now, we  state the main result of this Section:

\vs 3mm

\no{\bf Proposition 6.}  {\it There exists a constant $ \delta >0$,
$\delta =\delta(R, \rho)$ but independent
 of $T$, such that
 $\forall {\bf s_1}, {\bf s_2} \in C_s$ and $\forall
B \subset C_s  $,
\qq
\lambda^2 (B | {\bf s_1}) + \lambda^2 (B^c | {\bf s_2}) \geq \delta
\label{b0}
\qqq
}
\vspace*{5mm}

\no{\bf Remark.} The important point in this Proposition is that
$\delta$ is independent of $T$. The same will be true
about the constants $\delta_1$,  $\delta_2$, used in the proof (see
(\ref{a22}), (\ref{a23})).
\vs 5mm

Before proving this Proposition, we use it to give the
\vspace*{5mm}

\no{\bf Proof of Proposition 4.}
We shall use the previous Proposition
 and a slightly modified version of an argument taken from \cite{Doob}, p.
197--198. Let, for $B
\subset C_s$,
$$
{\underline \lambda}(n, B)=\inf_{{\bf s}\in
C_s}\lambda^n(B|{\bf s}),\;\;\;\; {\overline \lambda}(n,
B)=\sup_{{\bf s}\in C_s}\lambda^n(B|{\bf s}).
$$
Fix ${\bf s_1}, {\bf s_2}\in C_s$ and consider the
function defined on subsets $B\subset C_s$:
$$
\psi_{{\bf s_1},{\bf s_2}}(B)=\lambda^2(B|{\bf s_1})-\lambda^2(B|{\bf s_2}).
$$
Let $S^+$ be the set such that $\psi_{{\bf s_1},{\bf
s_2}}(B)\geq 0$ for $B\subset S^+$ and $\psi_{{\bf
s_1},{\bf s_2}}(B)\leq 0$ for $B\subset C_s\backslash
S^+\equiv S^-$ ($S^\pm$ depend on ${\bf s_1},{\bf s_2}$,
but we suppress this dependence). Observe that, by  (\ref{p2}, \ref{p3}), we have,
$\forall {\bf s}
\in C_s$,
\qq
1-e^{-cT} \leq \lambda^2 (C_s|{\bf s}) \leq 1.
\label{PTQ}
\qqq
(with a smaller $c$ than in (\ref{p2}, \ref{p3})).
 Then,
\qq
|\psi_{{\bf s_1},{\bf s_2}}(S^+)+\psi_{{\bf s_1},{\bf
s_2}}(S^-)|=|\lambda^2(C_s|{\bf s_1})-\lambda^2(C_s| {\bf
s_2})|
\leq  e^{-cT}.
\label{psio}
\qqq
 Moreover, using (\ref{PTQ}, \ref{b0}), and $S^+\cup S^- = C_s$,
\qq
&&\psi_{{\bf s_1},{\bf s_2}}(S^+)=\lambda^2(S^+ |{\bf
s_1})-\lambda^2(S^+ |{\bf s_2})\nonumber\\ &&\leq
1-(\lambda^2(S^- |{\bf s_1})+\lambda^2(S^+ |{\bf
s_2}))\leq 1-\delta.
\label{psi1}
\qqq
Thus,
\qq
{\overline \lambda}(t+2, B)-{\underline \lambda}(t+2,
B)&=&\sup_{{\bf s_1},{\bf s_2}}\int ( \lambda^2(d{\bf
s}|{\bf s_1})-\lambda^2(d{\bf s}|{\bf s_2}))
\lambda^t({B|\bf s})\nonumber\\ &=&\sup_{{\bf s_1},{\bf
s_2}}\int
\psi_{{\bf s_1},{\bf s_2}}(d{\bf s})\lambda^t(B| {\bf s})\nonumber\\
&\leq& \sup_{{\bf s_1},{\bf s_2}}(\psi_{{\bf s_1},{\bf
s_2}}(S^+){\overline \lambda}(t, B) +\psi_{{\bf s_1},{\bf
s_2}}(S^-){\underline \lambda}(t, B))\non\\&=&
\sup_{{\bf s_1},{\bf s_2}}
(\psi_{{\bf s_1},{\bf s_2}}(S^+)({\overline \lambda}(t,
B)-{\underline
\lambda}(t, B))+(\psi_{{\bf s_1},
{\bf s_2}}(S^+)+\psi_{{\bf s_1},{\bf s_2}}(S^-
)){\underline \lambda}(t, B))
\nonumber\\ &\leq&
(1-\delta)({\overline \lambda}(t, B)-{\underline
\lambda}(t, B))+e^{-cT}
\nonumber
\qqq
where, to get the last inequality, we used (\ref{psi1})
and (\ref{psio}) and ${\underline \lambda}(t, B)
\leq1$.
We conclude that, $\forall \Ns' \in C_s$,
$$
|\lambda^{2n}(B|{\bf s'})-\lambda^{2n}(B|0)|\leq
{\overline
\lambda}(2n, B)-{\underline \lambda}(2n, B)
\leq (1-\delta)^{n-1}
+\frac{e^{-cT}}{\delta}.
 $$
Now, choose first $n$
 sufficiently
large so that $(1-\delta)^{n-1}\leq \frac{1-{\bar
\delta}}{4}$, for
some ${\bar \delta}>0$ and then $T$ sufficiently large so
that $\frac{e^{-cT}}{\delta}\leq \frac{1-{\bar
\delta}}{4}$. Since, with $p=2n$,
$$\int|\lambda^p(d\Ns|\Ns')-\lambda^p(d\Ns|0)|
\leq 2\sup_B |\lambda^{2n}(B|{\bf s'})-\lambda^{2n}(B|0)|,
$$
(\ref{mun}) follows, with $\delta$ in that equation equal to
$\bar
\delta$ here.\hfill$\Box$

\vspace*{5mm}
\no{\bf Proof of Proposition 6.}
First of all, observe that it is enough to prove
(\ref{b0}) with $\lambda$ replaced by $P$:
\qq
P^2 (B | {\bf s_1}) + P^2 (B^c | {\bf s_2}) \geq \delta
\label{b00}
\qqq
 since we can then
use (\ref{p2}) and choose $T$ large enough to obtain the
same result for $\lambda$, since $\delta$ is independent
of $T$.

It will be convenient to write $P^2 (d{\bf s}_+|{\bf
s_1})
=\int P (d{\bf s}_+ |  {\bf s} )P (d {\bf s}
| {\bf s_1})$, where we write ${\bf s}_+ \in C^+_s$
meaning ${\bf s}_+\in C_s\subset C([0,T], H_s)$ (see
(\ref{Cs})), and similarly ${\bf s_1}
\in C^-_s \subset  C([-2T,-T], H_s)$, ${\bf s}
\in C^0_s \subset  C([-T,0],H_s)$, which is the variable
over which we integrate.

Turning to the proof, we first get a lower bound on
(\ref{b00}) by replacing $B$, $B^c$ by $B \cap V^+$, $B^c
\cap V^+$, where $V^+  $ is defined by
\qq
V^+ = \{{\bf s}_+  \in C^+_s | \sum_{n=1}^{t} D_{
n}(\omega_0) < \zeta Rt,
\;\; \forall t \in [1, T]
\}
\label{b2}
\qqq
where $\zeta$ will be chosen large enough below and
$\omega_0(t)=s(t)+l(t, {\bf s}_+([0, t]), 0)$.
To simplify the notation, we shall assume,
from now on, that $B
\subset V^+$ and $B^c
\equiv V^+ \backslash B$.

Next, we obtain also a lower bound
on
$P^2 (B | {\bf s_1}) = \int P (B|  {\bf s} )
P (d{\bf s}  |  {\bf s_1} )$ and on $P^2 (B^c | {\bf
s_2})$ by restricting
the integrations over ${\bf s}$, so that we have:
\qq
(\ref{b00}) \geq \int P (B | {\bf s})
 1 ({\bf s}|{\bf s_1}) P (d{\bf s} | {\bf s_1}) +
\int P (B^c | {\bf s}) 1 ({\bf s}|{\bf s_2})
 P (d{\bf s} | {\bf s_2})
\label{a17}
\qqq
where $1 ({\bf s}|{\bf s'}) =1_01_{[-1,0]} 1_{\leq -1}$
with
\qq
 &&1_0(s(0))=1(\|s(0)\|^2
\leq 3{\zeta'}R),\non\\ &&1_{[-1,0]} ({\bf s}([-1,0]))
= 1 (\displaystyle{\sup_{t\in [-1,0]}}
\|s(t)\|^2 \leq \zeta R),\non\\
&&1_{\leq -1}({\bf s}([-T,-1])|{\bf s'})=
 1(\|\omega(-1)\|^2 \leq {\zeta'}R),
\label{e01}
\qqq
where  $\omega(-1)=s(-1) +l(-1, {\bf s}([-T, -1]), l
(-T))$, with $l (-T)= l(-T,{\bf s'} ([-2T, -T]), 0 )$,
and $\zeta $, $\zeta'$ are constants that will be chosen
large enough below, but with $\zeta'
\leq C \zeta$ for $C$ large ($\zeta$, $\zeta'$
 play a role somewhat similar to $\beta$, $\beta'$
 in the previous Sections, but they are not necessarily equal
  to the latter).

Before proceeding further, let us explain the basic idea of the proof.
To prove (\ref{b00}), it would be enough to bound
${P^2(B|{\bf s_1})\over P^2 (B|{\bf s_2})} \geq \delta$.
We do not quite do that, but first give, in Lemma 8.1 below,
 a lower bound
on ${P(B|{\bf s})\over P (B|{\bf s}')}$ for ${\bf s},
{\bf s}'$ in a ``good" set of configurations, i.e. in the
support of the indicator functions that we just
introduced. Good here means that the ``interaction" (or,
to be more precise, the analogue of what is called in
Statistical Mechanics the relative Hamiltonian),
expressed through
 the Girsanov formula (see e.g.  (\ref{a24})), between
the paths in $C_s^0$ that are in the support of those
indicator functions and those in $V^+$ is, in some
sense, bounded. This relies on Lemma 8.5, which itself
follows from the results of the previous Section. Next,
we show that the probability of reaching that good set,
does not depend very much on whether we start from
 ${\bf s_1}$ or ${\bf s_2}$ in $ C_s^-$ (see Lemma 8.2).
This is rather straightforward, but depends on standard
 estimates on the Brownian bridge (see Lemmas 8.6 and 8.7)
that we give in detail, for the sake of completeness.
Finally, we need to show that the probability of the
set of good configurations, as well
as the one of $V^+$, is bounded from below; this is done in Lemma 8.3.
Remember that all the bounds here have to be $T$-independent, since this was
used in an essential way in the proof of the Theorem (Section 6).

Now, we shall state and use the Lemmas that we need and that will be
proven below. Let
$$
W =\cup_{{\bf\bar s}\in C_s^-}{\rm supp} (1(\cdot|
{\bf\bar s})).
$$

\vs 3mm

\noindent
{\bf Lemma 8.1.} {\it  $\exists c=c(R,\rho)>0$, such
that, $\forall B \subset V^+$,
$\forall {\bf s}, {\bf s}'
\in W$
with $s(0)= s'(0)$ and $P(B|{\bf s}') \neq 0$:
\qq
{P(B|{\bf s})\over P (B|{\bf s}')} \geq e^{-
{c(R,\rho)\over P(B|{\bf s}')}}
\label{a19}
\qqq
}
\vs 3mm

\no Defining
\qq
h(B,s_0) = \sup_{{\bf s}\in W:\, s(0)=s_0} P (B | {\bf
s})
\label{a18}
\qqq
we conclude from the Lemma that for all ${\bf s}
\in W$ such that $s(0) = s_0$,
\qq
P (B | {\bf s}) \geq \ell_B (s_0) \equiv h(B,s_0) e^{-
{c(R,\rho) \over h(B,s_0)}},
\label{a20}
\qqq
where both sides vanish if $h(B,s_0)=0$.
Hence, applying the same argument to $P (B^c|{\bf s})$, we get:
\qq
(\ref{a17}) \geq E ( \ell_B1 (\cdot|{\bf s_1}) | {\bf
s_1}) + E ( \ell_{B^c}1(\cdot|{\bf s_2}) | {\bf s_2})
\label{a21}
\qqq
where here $\ell_{B}$, $\ell_{B^c}$ are functions of
$s(0)$ and $E$ is the (conditional) expectation.

The next lemma controls the dependence on the past in
(\ref{a21}):

\vs 3mm

\noindent
{\bf Lemma 8.2.} {\it $\exists \delta_1 > 0$, $\delta_1=
\delta_1 (R, \rho)$ such that $\forall {\bf s_1}, {\bf
s_2} \in C_s^-$, $\forall B \subset V^+$,
\qq
{E (\ell_{B^c}1(\cdot|{\bf s_2}) | {\bf s_2}) \over
E(\ell_{B^c}1(\cdot|{\bf s_1}) | {\bf s_1})}
\geq
\delta_1.
\label{a22}
\qqq
provided that, in (\ref{b2}), (\ref{e01}),  $ \zeta' $ is
large enough and $ \zeta \geq C \zeta'$ for $C$ large. }
\vs 3mm
Then, since any $\delta_1 $ satisfying (\ref{a22}) must
be less than $1$,
\qq
(\ref{a21}) \geq \delta_1 (E ( \ell_B1(\cdot|{\bf s_1}) |
{\bf s_1}) + E ( \ell_{B^c}1(\cdot|{\bf s_1})  | {\bf
s_1})).
\label{a221}
\qqq
But, we also have:

\vs 3mm

\noindent
{\bf Lemma 8.3.}
{\it   $\exists \delta_2 > 0$, $\delta_2=
\delta_2 (R, \rho)$, such that, $\forall
 {\bf s'} \in C_s^-$,
\qq
\int P ( V^+ | {\bf s})1({\bf s}|{\bf s'})
P (d {\bf s}|{\bf s'}) \geq
\delta_2,
\label{a23}
\qqq
and
\qq
\int 1({\bf s}|{\bf s'})
P (d {\bf s}|{\bf s'}) \geq
\hf,
\label{x10}
\qqq
provided that, in (\ref{b2}), (\ref{e01}),  $ \zeta' $ is
large enough and $ \zeta \geq C \zeta'$ for $C$ large. }

\vs 2mm

\no{\bf Remark.} The important point here is that $\delta_2$
is independent of  $T$; to show this, we will use the
fact that, in (\ref{b2}),  the condition on $\sum_1^t$
increases sufficiently fast in time, so that, see below,
(\ref{a031}) is finite (however, it should not grow too
fast because, to prove Lemma 8.1, we need that it does
not grow faster than linearly, so that  (\ref{a27}) below
holds, leading to the finiteness of (\ref{a280})).

\vs 2mm

 By definition (\ref{a18}) of
$h_B$, and using this Lemma, we have
\qq
&&E (h_B1(\cdot|{\bf s_1})  | {\bf s_1})+E
(h_{B^c}1(\cdot|{\bf s_1})  | {\bf s_1})
\non\\ &&\geq \int P ( B | {\bf s})1({\bf s}| {\bf s_1}) P (d
{\bf s}|{\bf s_1}) +\int P ( B^c | {\bf s})1({\bf s}|
{\bf s_1}) P (d {\bf s}|{\bf s_1})\non\\ && = \int P (
V^+ | {\bf s})1({\bf s}| {\bf s_1}) P (d {\bf s}|{\bf
s_1}) \geq
\delta_2,
\label{h1}
\qqq
since $B \cup B^c = V^+$.
 Now, we need the following
straightforward consequence of Jensen's inequality:
\vs 3mm
\par\noindent
{\bf Lemma 8.4.} {\it For any probability measure $P$,
$$
E(\ell 1 ) \geq E (h1 ) \exp(-{cE(1)\over E(h1)}) \geq E
(h1 )
\exp(-{c \over E(h1)}),
$$
where $E$ is the expectation with respect to $P$, $\ell =
h e^{-{c
\over h}}$, the functions $h, 1$, satisfy $0\leq h$,
$0\leq 1
\leq 1$,  $h$ is integrable, and $c\in \NR_+$.
}
\vs 3mm

>From (\ref{h1}), we may assume $E (h_B1(\cdot|{\bf s_1})
| {\bf s_1})
\geq \frac{\delta_2}{2}$ (if not, exchange $B$ and $B^c$). Hence,
applying Lemma 8.4 to $E ( \ell_B1(\cdot|{\bf s_1}) |
{\bf s_1}) $, we get
$$
E ( \ell_B1(\cdot|{\bf s_1}) | {\bf s_1}) \geq
\frac{\delta_2}{2}
 \exp(-\frac{2c(R,\rho)}{\delta_2}).$$

So, combining this with (\ref{a17}), (\ref{a21}), (\ref{a221}),
we get :
\qq
(\ref{b00}) \geq  \frac{\delta_1 \delta_2}{2}
 \exp(-\frac{2c(R,\rho)}{\delta_2})
\non
\qqq
which finishes the proof of the Proposition.\hfill$\Box$
\vs 3mm

Now, we still have to prove Lemmas 8.1, 8.2, 8.3.

\vs 3mm
\noindent
{\bf Proof of Lemma 8.1.}
Recalling (\ref{Pss'}) we have
\qq
P(B | \Ns) = \int e^{\int^T_0 (f,\gamma^{-1} (ds_+ -
{1\over 2} f dt))} 1_B \nu_{s(0)} (d\Ns_+) \equiv \int g
1_B
 \nu_{s(0)}
(d\Ns_+)
\label{a24}
\qqq
where $f(t) = f (t,  \Ns_+([0, t]), l(0))$, with $l(0)
= l(0, \Ns([-T, 0]), 0 )$, is a function of $
\Ns_+ \vee \Ns$ (the symbol $\vee$ was defined after
equation (\ref{Pss'})), and $ \nu_{s(0)} (d\Ns_+)$ is the
Wiener measure with covariance $\gamma$, on paths
starting at $s(0)$. $P (B |
\Ns')$ is defined similarly with $f'(t)= f (t,  \Ns_+([0, t]), l'(0))$
and $l'(0)= l(0, \Ns'([-T, 0]), 0 )$. The corresponding
Girsanov factor is denoted $g'$. Since $s(0)=s'(0)$, we
can write
\qq
{P(B|\Ns)\over P(B|\Ns')} &&= Ee^{\int^T_0
(f,\gamma^{-1}(ds_+ - {1\over 2}
fdt))-(f',\gamma^{-1}(ds_+
- {1\over 2}f'dt))}\non\\
&&= E e^{\int^T_0 (f-f',\gamma^{-1} (ds_+
- f' dt)) - {1\over 2} \int (f-f', \gamma^{-1} (f-f'))dt}
\label{a25}
\qqq
where the expectation is taken with respect to the normalized
measure
\qq
{1_B g' \nu_{s(0)} (d\Ns_+) \over \int 1_B g'
\nu_{s(0)} (d\Ns_+)} = {1_B g' \nu_{s(0)} (d\Ns_+)
\over P(B | \Ns') } .
\label{a025}
\qqq
 By Jensen's inequality,
\qq
(\ref{a25}) \geq e^{E(\int^T_0 (f-f',\gamma^{-1} (ds_+
- f' dt)) - {1\over 2} \int (f-f', \gamma^{-1} (f-f'))dt)}
\label{a26}
\qqq

We will bound the argument of the exponential. For that,
we need some  estimates that follow from  the results
of  the previous Section:

\vs 3mm
\par\noindent
{\bf Lemma 8.5.} {\it $\forall \Ns, \Ns'
\in W$ and
 $\forall \Ns_+ \in V^+$,
\qq
\| f(t) - f' (t)\| \leq C (R) e^{-ctR}.
\label{a27}
\qqq
}
The proof of this Lemma will be given at the end of this Section.
Returning to the proof of Lemma 8.1,
\qq
|E \int^T_0 (f-f', \gamma^{-1} (f-f') )dt |
\leq
c(R,\rho).
\label{a28}
\qqq
since $\gamma_k\geq\rho $ and, by (\ref{a27}),
\qq
\int_0^\infty \| f(t) - f' (t)\|^2 dt \leq C(R).
\label{a280}
\qqq
To bound the stochastic integral in
(\ref{a26}) we proceed as in Section 7 by defining
$$
\eta(t) = 1 \Bigl(\| f(t) - f' (t)\| \leq C (R) e^{-ctR} \Bigr)
$$
with $c$, $C(R)$ as in (\ref{a27}).
 Since the measure
with respect to which the expectation $E$ is taken has
support in $B \subset V^+$ and since (\ref{a27}) holds in
$V^+$, we can write, see (\ref{a025}),
\qq
&&|E(\int^T_0 (f-f', \gamma^{-1} (ds_+ - f' dt))| = |\int
g' d \nu_{s(0)} (\int^T_0 (\eta (f-f'), \gamma^{-1} (ds_+
- f' dt)))1_B|  \non\\ &&\leq {(E_b (\int^T_0 (\eta
(f-f'),\gamma^{-1}db))^2)^{1\over 2} (\int g' d
\nu_{s(0)} 1^2_B)^{1\over 2} \over P(B|\Ns')}
\label{a281}
\qqq
where we changed variables: $ds_+ - f' dt =db$,
using Girsanov's formula (backwards),
 and
where $E_b$ denotes the expectation with
respect to Brownian motion with covariance $\gamma$.
Finally, using (\ref{a280}) on the support of $\eta$
and the fact that $\int g' d \nu_{s(0)} =1$, we get:
\qq
(\ref{a281})\leq {c(R,\rho) \over P (B |
\Ns')}.
\non
\qqq
Combining this, (\ref{a28}) and (\ref{a26}), we conclude
\qq
{P(B|\Ns)\over P(B|\Ns')}\geq e^{-\frac{c(R,\rho)}{P (B |
\Ns')}}.
\non
\qqq
which proves the Lemma.\hfill$\Box$

\vs 3mm

Let us turn to Lemma 8.2. It will be useful to study in
some detail the paths over the interval $[-1,0]$. Let
$\nu_{s_{-1}s_0} (d\Ns)$ be the (unnormalized) measure defined by the
Brownian bridge going from $s_{-1}$ at time $-1$ to $s_0$
at time 0, whose total mass is:
\qq
M(s_0, s_{-1}) = \prod_k {1 \over 2\pi
\gamma_k} \exp\Bigl(-{|s_{0k}-s_{-1k}|^2 \over
2\gamma_k}\Bigr),
\label{a39}
\qqq
where the product runs over $k$ such that $|k|^2 \leq
\kappa R $. Define
\qq
P(s_0, s_{-1} | \Ns \vee {\bf s_1}) = \int e^{\int^0_{-1}
(f,\gamma^{-1} (ds (t) - \hf fdt))} 1_{[-1,0]} (\Ns )
\nu_{s_{-1}s_0} (d\Ns).
\label{a35}
\qqq
where $f(t)= f(t, \Ns[-1, t], l(-1))$, with $l(-1)
=l(-1, \Ns \vee {\bf s_1} ([-2T, -1]), 0)$ and
similarly
\qq
P_{s_{-1}}(d{\bf s}|{\bf s_1})=
e^{\int_{-T}^{-1} (f,\gamma^{-1} (ds (t) - \hf fdt))}
 \nu_{s_{1}(-T) s_{-1}} (d\Ns).
\label{y1}
\qqq
Then we can write
\qq
E (1 (\cdot|{\bf s_1})\ell_{B^c} |{\bf s_1}) =
\int \ell_{B^c} (s_0)  1_0(s_0) P (s_0,s_{-1} | \Ns \vee {\bf s_1})
 1_{\leq -1}({\bf s}|{\bf s_1})P_{s_{-1}}(d{\bf s}|{\bf s_1})
 ds_0 ds_{-1}.
\label{a035}
\qqq
We shall need
\vs 3mm
\par\noindent
{\bf Lemma 8.6.} {\it $\exists C_1, C_2$, $ C_i= C_i
(R,\rho)$, $i=1,2$, such that $\forall {\bf \bar s} \in
C_s^-$, and $\forall s_0, s_{-1},
\Ns \in {\rm supp}(1(\cdot| {\bf \bar s})) $:
\qq
C_1\leq P(s_0, s_{-1} | \Ns \vee {\bf \bar s})\leq C_2.
\label{a33}
\qqq
provided that, in (\ref{b2}), (\ref{e01}),  $ \zeta' $ is
large enough and $ \zeta \geq C \zeta'$ for $C$ large. }
\vs 3mm
>From this, Lemma 8.2 follows easily:

\vs 3mm

\noindent{\bf Proof of Lemma 8.2.} Using
(\ref{a035}), we have:
\qq
{E ( \ell_{B^c}1(\cdot|{\bf s_2}) |{\bf s_2})\over E(
\ell_{B^c}1(\cdot|{\bf s_1}) | {\bf s_1})} &&=\frac{\int
\ell_{B^c} (s_0) 1_0(s_0) P (s_0,s_{-1} |
\Ns \vee {\bf s_2})1_{\leq - 1}({\bf s}|{\bf
s_2}) P_{s_{-1}} (d\Ns |{\bf s_2}) ds_0 ds_{-1}}{\int
\ell_{B^c} (s_0) 1_0(s_0) P (s_0,s_{-1} | \Ns \vee {\bf
s_1})1_{\leq - 1}({\bf s}|{\bf s_1}) P_{s_{-1}} (d\Ns
|{\bf s_1}) ds_0 ds_{-1}}
\non \\
&&\geq \inf_{s_0} { \int  P(s_0,s_{-1} | \Ns \vee {\bf
s_2})1_{\leq -1}({\bf s}|{\bf s_2}) P_{s_{-1}} (d\Ns |
{\bf s_2})ds_{-1}\over  \int P(s_0,s_{-1} | \Ns \vee {\bf
s_1}) 1_{\leq -1}({\bf s}|{\bf s_1})P_{s_{-1}} (d\Ns|{\bf
s_1})ds_{-1}}
\label{a32}
\qqq
where the infimum is taken over $s_0 \in {\rm supp}
(1_0)$. Now use (\ref{a33}) and
$$
\int
1_{\leq -1}({\bf s}| {\bf s_i}) P_{s_{-1}} (d\Ns|{\bf
s_i})ds_{-1}
=\int 1_{\leq -1}({\bf s}| {\bf s_i}) P (d\Ns|{\bf s_i}),
$$
for $i=1,2$ to bound from below (\ref{a32}) by
\qq
(\ref{a32}) \geq \frac{C_1 \int 1_{\leq -1}({\bf s}|{\bf
s_2}) P (d\Ns | {\bf s_2})} {C_2\int 1_{\leq -1}({\bf
s}|{\bf s_1}) P (d\Ns |{\bf s_1})} \geq
\frac{C_1}{C_2}\int 1_{\leq -1}({\bf s}|{\bf
s_2})P (d\Ns | {\bf s_2}) \geq  \frac{C_1}{2C_2} =\delta_1.
\non
\qqq
where in the last inequality, we used:
\qq
\int1_{\leq -1}({\bf s}| {\bf s_2}) P (d\Ns | {\bf s_2})
\geq \int 1({\bf s}| {\bf s_2})
P (d {\bf s}|{\bf s_2})  \geq {1\over 2},
\label{a34}
\qqq
where the first inequality is trivial, see (\ref{e01}),
and the second follows from (\ref{x10}) in  Lemma
8.3.\hfill$\Box$
\vs 3mm
Now, we will prove Lemma 8.6, to complete the proof of Lemma 8.2,
before proving Lemma 8.3.

\vs 3mm

\noindent
{\bf Proof of Lemma 8.6.}  We write, for $t \in [-1,0]$ :
\qq
s(t) = (1+t) s_0 - t s_{-1} + \alpha (t)
\label{a36}
\qqq
where $\alpha(\cdot)$ is the  Brownian bridge with
covariance $\gamma$, going from $0$ at time $-1$ to $0$ at
time $0$, i.e. the Gaussian process with covariance:
\qq
E(\alpha_k (t') \alpha_p (t)) =
\delta_{k,-p}
\gamma_k(1+t')(-t) \hspace{5mm} (-1 \leq t' \leq t \leq 0)
\label{a37}
\qqq
for $k^2, p^2 \leq \kappa R$.
Substituting  (\ref{a36})  into (\ref{a35}), we get:
\qq
P(s_0,s_{-1}|\Ns \vee {\bf \bar s}) = M (s_0,s_{-1})\int
e^{
\int^0_{-1} (f,\gamma^{-1}(d\alpha (t)+(s_0 - s_{-1}-\hf f) dt))}
1_{[-1,0]}({\bf s}) \nu (d{\bf \alpha})
\label{a38}
\qqq
where $\nu$ is the probability distribution of
 the Brownian bridge $\alpha$.

To bound $ M (s_0, s_{-1})$ remember, from (\ref{e01}),
that, for $s_0$, $s_{-1}$ in the support of $1(\cdot|{\bf
\bar s})$, we have
\qq
\|s_0 \|^2 \leq 3{\zeta'} R,
\label{038}
\qqq
 and
\qq
\|s_{-1}\|^2 \leq \|
\omega (-1) \|^2 \leq {\zeta'} R.
\label{039}
\qqq
These bounds, combined with the definition (\ref{a39}) of
$M(s_0,s_{-1})$ imply that, for $s_0$, $s_{-1}$ in the
support of $1(\cdot|{\bf \bar s})$,
\qq
C_2 (R, \rho) \leq M(s_0,s_{-1}) \leq C_1 (R, \rho).
\label{a40}
\qqq
Thus, to prove (\ref{a33}), we need only to bound from
above and from below the integral
\qq
\int
e^{\int^0_{-1} (f,\gamma^{-1}(d\alpha(t) + (s_0 - s_{-1}-\hf
f) dt))} 1_{[-1,0]} ({\bf s}) \nu (d{\bf \alpha})
\label{040}
\qqq
by a constant depending only on $R$ and $\rho$.
For this, some elementary facts about Brownian bridge
will be needed:

\vs 3mm

\noindent
{\bf Lemma 8.7.}
{\it Let $\alpha$ be the Brownian Bridge on $[-1,0]$ with
covariance $\gamma$.  Then

\no $(a)$. There exists a constant $c(R)>0$ such that
\qq
\int 1( \sup_{\tau\in [-1,0]} \| \alpha (\tau) \|^2 \leq {\zeta'} R)
\nu (d{\bf \alpha})
\geq c(R).
\label{a042}
\qqq

\no $(b)$.  Let $g(t)$ be progressively measurable with
${\sup_{t\in[-1,0]}}
\|g(t)\|
\leq A$.  Then
\qq
\int e^{\int^0_{-1} (g,d\alpha)} \nu (d{\bf \alpha}) \leq C(A,R, \rho),
\label{a45}
\qqq
and
\qq
\int (\int^0_{-1} (g, d\alpha))^2 \nu(d{\bf \alpha}) \leq C(A,R,\rho)
\label{a46}
\qqq
}
\vs 3mm

Continuing with (\ref{040}), we need some bounds on
$\|f(t)\| $ for $\Ns$ in the support of $1(\cdot|{\bf
\bar s})$. First, we have, $\forall
\Ns
\in {\rm
supp} (1(\cdot|{\bf \bar s}))$,
\qq
\sup_{t\in [-1,0]} \| l(t)\| \leq C (R)
\label{a027}
\qqq
where $l(t) = l(t, {\bf s} ([-1, t]), l (-1))$,
 which holds combining (\ref{l10})
in Proposition 1,
 and the fact that, on ${\rm
supp} (1(\cdot|{\bf \bar s}))$ (see (\ref{e01})), both
$\omega (-1)$ and $\sup_{t\in [-1, 0]} \|s(t)\|^2$ are of
order $R$. This and $\sup_{t\in [-1, 0]} \|s(t)\|^2\leq
\zeta R$ on ${\rm supp}
(1(\cdot|{\bf \bar s}))$ imply that $\| \omega(t)\|$ also
satisfies  (\ref{a027}). Then, using (\ref{a14}), we
get:
\qq
\sup_{t\in [-1, 0]}\|f(t)\| \leq C(R).
\label{041}
\qqq

Consider now the lower bound on (\ref{040}).  By Jensen's
inequality,
\qq
(\ref{040}) \geq C (R, \rho) \left[\int 1_{[-1,0]}({\bf
s}) \nu (d{\bf
\alpha})\right]
\exp
\left[\frac{\int (\int^0_{-1}
(f,\gamma^{-1}d\alpha)1_{[-1,0]}({\bf s})\nu(d{
\alpha})}{\int
1_{[-1,0]}({\bf s})
\nu (d\alpha)}\right]
\label{a41}
\qqq
where $C(R, \rho)$ is a lower bound on $\exp(\int^0_{-1}
(f,\gamma^{-1} (s_0 - s_{-1}-\hf f )dt )$ (which holds  because of
(\ref{041}) and (\ref{038}), (\ref{039})).

Using
(\ref{038}), (\ref{039}),
 we obtain from
(\ref{a36})
 that, for $s_0$, $s_{-1}$ in the support
 of $1(\cdot|{\bf \bar s})$, if $\sup_{t\in [-1,0]} \| \alpha (t) \|^2
  \leq {\zeta'} R$,
 then $\sup_{t\in [-1,0]} \| s(t) \|^2
 \leq C{\zeta'} R$
 for $\zeta \geq C\zeta'$; hence,
\qq
1_{[-1,0]}({\bf s}) \geq 1( \sup_{\tau\in [-1,0]} \|
\alpha (\tau) \|^2 \leq {\zeta'} R) .
\non
\qqq

Combining this with (\ref{a41}) and (\ref{a042}), we get
\qq
(\ref{040})\geq c(R)C(R, \rho)e^{-c(R)^{-1}{\int
|\int^0_{-1}(f,\gamma^{-1}d\alpha) |1_{[-1,0]}({\bf s})
\nu (d{\bf \alpha})}}
\label{a43}
\qqq
Now, let $g=f\gamma^{-1}$ and use Schwarz'
inequality to get the upper bound
\qq
\int |\int^0_{-1} (f,\gamma^{-1}d\alpha)|1_{[-1,0]}({\bf s}) \nu
(d{ \alpha})\leq \Bigl(\int (\int^0_{-1} (g, d\alpha))^2 \nu(d{\bf
\alpha})\Bigr)^\hf,
\label{s10}
\qqq
(since $\int \nu (d\alpha) =1$). Using
 (\ref{041}) and $\gamma_k \geq \rho$ we obtain that
$\displaystyle{\sup_{t\in[-1,0]}} \|g(t)\|
\leq A=C(R, \rho)$
and so (\ref{a46}) leads to  an upper bound  $C(R,\rho)$
for (\ref{s10}) and thus, a lower bound $C(R,\rho)$ for
(\ref{040}).

 Finally,
we bound from above   (\ref{040}) by $C(R,\rho)$, using
$1_{[-1, 0]}({\bf s})\leq 1$ and then combining
 (\ref{041}),
(\ref{038}), (\ref{039}),
 and
 (\ref{a45}) with $g=f\gamma^{-1}$.\hfill$\Box$

\vs 3mm

\noindent
{\bf Proof of Lemma 8.7.} (a). Observe that $\alpha
(\tau)$ has the same distribution as $(-\tau)
b(-\frac{1+\tau}{\tau})$, where $b(\cdot)$ is the
Brownian motion starting at $0$, with covariance
$\gamma$. So that, with $t=-\frac{1+\tau}{\tau}$,
(\ref{a042}) translates into:
\qq
\int 1 \Bigl(\sup_{t\in [0,\infty[} (\frac {\| b(t) \|}{1+t})^2 \leq {\zeta'} R
\Bigr)
 \nu_0 (db)\geq c(R).
\non
\qqq
This is readily proven, since $b(t)\in \NR^d$ with
$d=d(R)$ and the covariance $\gamma$ has an $R$-dependent
upper bound.

(b). Let
$$M_t =
e^{2\int_{-1}^t(g,d\alpha-({\alpha\over
\tau}+\gamma g)d\tau)}
$$
It is easy to see that $M_t$ is a martingale (see e.g. \cite{simon} p. 158), and
that, therefore, $\forall t \in [-1, 0]$, $E(M_t) = 1$ where $E$ is
the expectation with respect to $\nu (d\alpha)$. So,
write
$$e^{\int^0_{-1}
(g,d\alpha)}=M_0^\hf e^{
\int_{-1}^0(g,({\alpha\over
\tau}+\gamma g))d\tau}
$$
 and use Schwarz' inequality and $E(M_0) = 1$ to get
\qq
\int e^{\int^0_{-1} (g,d\alpha)}\nu(d{\bf
\alpha})
\leq
(\int
e^{2
\int_{-1}^0(g,({\alpha\over
\tau}+\gamma g))d\tau})^\hf
\leq C(A,
R,\rho)
(\int e^{2\int^0_{-1}
{(g,\alpha)\over \tau}d\tau}
\nu
(d{\bf \alpha}))^\hf
\non
\qqq
where $C(A, R,\rho)$ is an upper bound on
$\exp \Bigl(\int^0_{-1} (g,\gamma g)d\tau \Bigr)$. Applying Jensen's
inequality to $e^{2\int^0_{-1}
{(g,\alpha)\over \tau}d\tau}$, with
$\frac{d\tau}{2\sqrt{|\tau|}}$ as probability measure on $[-1, 0]$,
we may bound the RHS by
\qq
C(A, R,\rho)(\int^0_{-1} {_{d\tau}\over^{2\sqrt{|\tau|}}}
\int e^{4A\|\alpha(\tau)\||\tau|^{-\hf}}\nu (d{\bf
\alpha}))^\hf,
\non
\qqq
where $ \displaystyle{\sup_{\tau \in [-1, 0]}}
\|g(\tau)\|
\leq A$ was used. To finish the proof, observe that
$$
\int e^{{4A\|\alpha(\tau)\||\tau|^{-\hf}}}\nu (d{\bf \alpha})\leq C(A, R, \rho)
$$
since $\|\alpha\| = (\sum_k |\alpha_k|^2)^{\hf} \leq
\sum_k |\alpha_k| $, and $\alpha_k(\tau)$ is a Gaussian
random variable with variance (see (\ref{a37})) $\gamma_k
(1+\tau)(-\tau)$.   (\ref{a46}) is an easy consequence of
(\ref{a45}).
\hfill$\Box$

\vs 3mm

This completes the proof of Lemma 8.6, hence of Lemma 8.2; so, we turn to the

\vs 3mm

\noindent {\bf Proof of Lemma 8.3.}
First, writing
\qq
\int P ( V^+ | {\bf s})1({\bf s}|{\bf s'})
P (d {\bf s}|{\bf s'})= \int ds_0
 P ( V^+ | (s_0, l (0,\Ns))1({\bf s}|{\bf s'})
e^{\int^0_{-T} (f,\gamma^{-1} (ds - {1\over
2} f dt))}
\nu_{{s'}(-T) s_0} (d {\bf s})
\label{x2}
\qqq
where $l(0,\Ns) = l(0, \Ns([-T, 0]), 0 )$ and $f(t)=
f(t,{\bf s}\vee {\bf s'}([-2T, t]), 0) $, we obtain the
lower bound:
\qq
\int P ( V^+ | {\bf s})1({\bf s}|{\bf s'})
P (d {\bf s}|{\bf s'})\geq I_1 I_2\int 1({\bf s}|{\bf
s'}) P (d {\bf s}|{\bf s'})
\label{x3}
\qqq
where
\qq
I_1= \inf_{{\bf s}} \frac{ P ( V^+ | (s_0, l(0,\Ns))}{ P
( V^+ | (s_0, 0))},
\label{x4}
\qqq
\qq
I_2 = \inf_{s(0)}  P ( V^+ | (s(0), 0))
\label{x5}
\qqq
and the infimum in (\ref{x4}) is taken over ${\bf s}\in
{\rm supp} (1(\cdot|{\bf s'}))$ with $s(0)=s_0$, while in
(\ref{x5}) it is taken over $s(0)\in {\rm supp}( 1_0)$ .
Now,  Lemma 8.1 implies that
\qq
I_1 \geq \exp \Bigl(- {c(R,\rho)\over I_2}\Bigr),
\label{x6}
\qqq
provided $I_2\neq 0$, which we shall show now. Since
$\omega_0(t)$ in terms of which $V^+$ was defined (see
(\ref{b2})) satisfies $\omega_0(0) =(s (0), 0)$, we can
write:
\qq
I_2 =1-\sup_{s(0)}( E (1-1_{ V^+} |\omega_0(0) )
\label{x7}
\qqq
To bound $E (1-1_{ V^+} |\omega_0(0) )$ we use the
probabilistic estimates (\ref{x1}):
\qq
P\Bigl(\sum^{t}_{n=1} D_n (\omega)> {\zeta R t}|
\omega(0) \Bigr)
\leq Ce^{-c\zeta t}
\label{a31}
\qqq
which holds for any $t$, $1\leq t\leq T$, and any
$\omega(0)$ with $\|\omega(0)\|^2\leq 3\zeta'R$, provided
$\zeta$ is large enough.
 Note that
this condition on
$\omega(0)$  holds for $\omega(0)= \omega_0 (0)=(s(0), 0)$ and
 $s(0) \in {\rm supp}(1_0)$
(see (\ref{e01})).

Thus, since $1-1_{ V^+}$ is the indicator function of the
event that
\qq
\sum_{n=1}^{t}D_{n}(\omega_0)\geq {\zeta Rt}
\label{h3}
\qqq
 for some $t \geq 1$, (\ref{a31}) applied to $\omega_0$ implies
\qq
 E (1-1_{ V^+} |\omega_0(0) ) \leq
\sum^\infty_{t=1} Ce^{-c\zeta t} \leq C e^{-c\zeta }.
\label{a031}
\qqq
and, by (\ref{x7}),
\qq
 I_2\geq
1-C e^{-c\zeta }.
\label{x9}
\qqq
This and (\ref{x6}) implies:
\qq
I_1 \geq \exp (-
c'(R,\rho)).
\label{x61}
\qqq

Finally, consider the last factor in (\ref{x3}); let us write
\qq
\int 1({\bf s}|{\bf s'})
P (d {\bf s}|{\bf s'}) = 1 - E((1-1(\cdot|{\bf s'})) |
{\bf s'}),
\label{x8}
\qqq
and let us bound from above  $E((1-1(\cdot|{\bf s'})) |
{\bf s'})$;
 remember that,
by (\ref{e01}), $1(\cdot|{\bf s'}))
=1_01_{[-1,0]} 1_{\leq
-1}$. We have
\qq
&&1-1(\cdot|{\bf s'}))=1-1_{\leq
-1} + (1-1_0 1_{[-1,0]}) 1_{\leq
-1}\non\\
&&\leq 1-1_{\leq
-1} + 1\Bigl(\sup_{t\in [-1,0]}
\|s(t)\|^2 \geq 3{\zeta'}R\Bigr) 1_{\leq
-1}.
\label{l3}
\qqq
where we
bounded $\zeta R \geq 3{\zeta'}R$, in the argument of
$1_{[-1,0]}$; So,
\qq
&&  E\Bigl((1-1(\cdot|{\bf s'})) \Big| {\bf
s'}\Bigr)\non\\ &&\leq
 E\Bigl(1(\|\omega(-1)\|^2
\geq {\zeta'}R )\Big| {\bf s'}\Bigr) +\sup  E\Bigl(
1(\displaystyle{\sup_{t\in [-1,0]}}
\|s(t)\|^2 \geq 3{\zeta'}R)\Big|\omega(-1)\Bigr)\non\\
&&=E\Bigl(1(\|\omega (-1)\|^2 >{\zeta'}R) \Big| \omega'
(-T)\Bigr) +\sup  E\Bigl( 1(\displaystyle{\sup_{t\in
[-1,0]}}
\|s(t)\|^2 \geq 3{\zeta'}R)\Big|\omega(-1)\Bigr),
\label{a231}
\qqq
where the last term comes from
$$E\Bigl(1(\displaystyle{\sup_{t\in [-1,0]}}
\|s(t)\|^2 \geq 3{\zeta'}R) 1_{\leq
-1}(\cdot|{\bf s'})  \Big| {\bf s'}\Bigr)\leq\sup
E\Bigl(1(\displaystyle{\sup_{t\in [-1,0]}}
\|s(t)\|^2 \geq 3{\zeta'}R)\Big| \omega(-1)\Bigr),
$$
and  the supremum is taken over all ${\bf s} \in {\rm
supp} (1_{\leq
-1}(\cdot|{\bf s'}))$, i.e. so that $\omega(-1)$ satisfies
$\|\omega(-1)\|^2 \leq {\zeta'}R$.

 The first term of
(\ref{a231}) is bounded by
\qq
E\Bigl(1(\|\omega (-1)\|^2 >{\zeta'}R) \Big| \omega'
(-T)\Bigr)
\leq C\exp(-c{\zeta'}),
\label{a329}
\qqq
for $T$ large: this follows from (\ref{c5}), with $0$
replaced by $-T$, t by $-1$ and the fact that, since
${\bf s'} \in C_s$, $\omega'(-T) $ satisfies, by
(\ref{Cs}),
\qq
 \| \omega' (-T) \| \leq 4 \beta' R T.
\label{a30}
\qqq

For the second term of (\ref{a231}), we use $D_t (\omega)
\geq \hf \|s(t)\|^2 $, $\|\omega(-1)\|^2 \leq {\zeta'}R$
 and (\ref{a1})
to bound it also by $C\exp(-c
{\zeta'})$. So, we have
\qq
E\Bigl((1-1(\cdot|{\bf s'})) \Big| {\bf s'}\Bigr)\leq
C\exp(-c{\zeta'}).
\label{l1}
\qqq

So, combining (\ref{x3}), (\ref{x61}),
 (\ref{x9})  and (\ref{x8}, \ref{l1}), we get that the LHS of
$$
(\ref{a23}) \geq \exp (- c'(R,\rho)) (1-
C\exp(-c{\zeta})) (1-C\exp(-c{\zeta'}))=\delta_2 >0
$$
for  $\zeta $, $\zeta'$  large enough; obviously
(\ref{x10}) folows from (\ref{x8}, \ref{l1}), for
$\zeta'$  large enough; this proves the
Lemma.\hfill$\Box$

\vs 2mm

We are left with the

\vs 2mm

\noindent
{\bf Proof of Lemma 8.5.} To prove (\ref{a27}),
 bound its LHS by
\qq
\|f(t) - f_0(t)\| +\|f_0(t) - f'(t)\|
\label{l21}
\qqq
where $f_0(t) = f(t, {\bf s}_+ ([0, t]), 0)$ corresponds to $\omega_0$.
Now, to bound each term in (\ref{l21}) by $C(R)e^{-cRt}$,
use (\ref{a13}), with $\omega$ there replaced by $\omega_0$ here,
to get:
\qq
\|f(t) - f_0(t)\| \leq C(R) ( \|\omega_0(t)\|\|\delta l(t)\|
 + \|\delta l(t)\|^2)
\label{l23}
\qqq
with $\delta l(t)= \omega(t) -\omega_0(t)$. We have, for
$t\geq 1$, the bound:
\qq
\|\delta l(t)\|\leq
\exp (-cRt)\|\delta l(0)\| =\exp (-cRt)\|l(0)\|
\label{l22}
\qqq
where the  equality holds since $\omega_0 (0)= (s(0), 0)$,
and the inequality follows from (\ref{21}) (with
$\omega_1$  replaced by $\omega_0$) and using the bound,
which holds  for $t\geq 1$ and where $[t]$
is the integer part of $t$:
$$a\int_0^t \|\nabla \omega_0\|^2
\leq a\sum_{n=1}^{[t]+1} D_n(\omega_0)\leq a\zeta R([t]+1)\leq \frac{\kappa R t}{2}
$$
for $t\geq 1$ and $\kappa$ large. For  $t\leq 1$,
(\ref{21}) yields: $\|\delta l(t)\|\leq C(R)\|\delta
l(0)\|=C(R)\| l(0)\|$, since, by definition (\ref{b2}) of
$V^+$, $\int_0^t \|\nabla \omega_0\|^2\leq
D_1(\omega_0)\leq \zeta R$. Finally, $\|\omega_0 (t)\|$
in (\ref{l23}) is bounded by $\|\omega_0 (t)\|^2\leq
D_{[t]+1}(\omega_0)\leq
\zeta R([t]+1)$,
which also follows  from the definition of $V^+$ and
which we can write as $\zeta R([t]+1) \leq C(R)\exp
(\frac{cRt}{2})$. Combining this with (\ref{l22},
\ref{l23}) gives
\qq
\|f(t) - f_0(t)\| \leq C(R) \exp (-\frac{cRt}{2})( \| l(0)\|
 + \|l(0)\|^2),
\label{x11}
\qqq
and a similar bound on $\|f_0(t) - f'(t)\|$ with $ l(0)$
replaced by $ l'(0)$. Now, on the support of $1(\cdot|
{\bf \bar s})$, for any $ {\bf \bar s}\in C_s^-$, i.e. in
$W$,  we have $\| l(0)\|\leq C(R)$, $\| l'(0)\|\leq C(R)$
(see (\ref{a027})), which finishes the proof of
(\ref{a27}).\hfill$\Box$


\begin{thebibliography}{90}

\bibitem{bkl1} J. Bricmont, A. Kupiainen, R. Lefevere:
{\it Probabilistic estimates
for the two dimensional stochastic Navier-Stokes equations}.
 J. Stat. Phys. {\bf 100} (3/4), (2000), 743-756.
(http://mpej.unige.ch/mp-arc/e/99-486).


\bibitem{bkl2}  J. Bricmont, A. Kupiainen, R. Lefevere:
{\it Ergodicity of the 2D Navier-Stokes Equations with random forcing.}
Preprint (http://mpej.unige.ch/mp-arc/e/00-264).



\bibitem{Doob} J.L. Doob: {\it  Stochastic Processes},  John Wiley,
New-York (1953).

\bibitem{ems} W. E, J.C. Mattingly, Ya.G. Sinai; private communication.

\bibitem{fm} F. Flandoli, B. Maslowski: {\it Ergodicity of the 2-D
Navier-Stokes equation under random perturbations.}
Commun. Math. Phys. {\bf  171} (1995), 119-141.

\bibitem{frisch} U. Frisch: {\it Turbulence}, Cambridge University
Press (1995).

\bibitem{ks} S. Kuksin, A. Shirikyan: {\it Stochastic dissipative PDE's
and
Gibbs measures}. Preprint, to appear in Commun. Math.
Phys.


\bibitem{mat} J. C. Mattingly: {\it Ergodicity of 2D Navier-Stokes
equations
with random forcing and large viscosity.}
Commun. Math. Phys. {\bf  206} (1999), 273-288.


\bibitem{simon} B. Simon: {\it Functional Integration and Quantum
Physics.}
 New York Academic Press (1979).




\bibitem{vf} M. I., Vishik, A.V. Fursikov: {\it Mathematical Problems of
Statistical Hydrodynamics.}
 Kluwer (1980).

\end{thebibliography}
\end{document}